\documentclass[pdftex,5p,twocolumn]{elsarticle}
\usepackage{txfonts}
\usepackage{natbib}
\bibpunct{(}{)}{;}{a}{}{,}
\usepackage{indentfirst} 
\usepackage{amssymb}
\usepackage{array}
\usepackage{verbatim}

\journal{Icarus}

\begin{document}

\begin{frontmatter}

\title{On the aerodynamic redistribution of chondrite components in protoplanetary disks}

\author[lmcm]{Emmanuel Jacquet\corref{cor1}}
\ead{ejacquet@mnhn.fr}
\author[lmcm]{Matthieu Gounelle}
\author[cea]{S\'{e}bastien Fromang}

\cortext[cor1]{Corresponding author}

\address[lmcm]{Laboratoire de Min\'{e}ralogie et de Cosmochimie du Mus\'{e}um, CNRS \& Mus\'{e}um National d'Histoire Naturelle, UMR 7202, 57 rue Cuvier, 75005 Paris, France.}
\address[cea]{CEA, Irfu, SAp, Centre de Saclay, F-91191 Gif-sur-Yvette, France  \& UMR AIM, CEA-CNRS-Univ. Paris VII, Centre de Saclay, F-91191 Gif-sur-Yvette, France.}



\begin{abstract}
Despite being all roughly of solar composition, primitive meteorites (chondrites) present a diversity in their chemical, isotopic and petrographic properties, and in particular a first-order dichotomy between carbonaceous and non-carbonaceous chondrites. We investigate here analytically the dynamics of their components (chondrules, refractory inclusions, metal/sulfide and matrix grains) in protoplanetary disks prior to their incorporation in chondrite parent bodies. 
We find the dynamics of the solids, subject to gas drag, to be essentially controlled by the ``gas-solid decoupling parameter'' $S\equiv \textrm{St}/\alpha$, the ratio of the dimensionless stopping time to the turbulence parameter. The decoupling of the solid particles relative to the gas is significant when $S$ exceeds unity. $S$ is expected to increase with time and heliocentric distance. On the basis of (i) abundance of refractory inclusions (ii) proportion of matrix (iii) lithophile element abundances and (iv) oxygen isotopic composition of chondrules, we propose that non-matrix chondritic components had $S<1$ when carbonaceous chondrites accreted and $S>1$ when the other chondrites accreted. This suggests that accretion of carbonaceous chondrites predated on average that of the other chondrites and that refractory inclusions are genetically related to their host carbonaceous chondrites.
\end{abstract}




\end{frontmatter}

\section{Introduction}
\label{Introduction}

  Meteorites are the witnesses of the first Myr of the solar system. Among these, \textit{chondrites} have undergone little change since their parent body (asteroid or comet) accreted, and thus preserve various solids formed in the protoplanetary disk: Most abundant in chondrites are \textit{chondrules}, millimeter-sized silicate spherules that formed through rapid melting and cooling of precursor material, via an as yet elusive mechanism \citep{ConnollyDesch2004}. \textit{Metal} and \textit{sulfide} grains are probably genetically related to chondrules \citep{Campbelletal2005}. \textit{Refractory inclusions}, in particular \textit{Calcium-Aluminum-rich Inclusions} (CAI), likely formed by gas-solid condensation at high temperatures \citep{Grossman2010}, probably in a single reservoir located close to the Sun \citep{MacPherson2005}---the ``CAI factory'' of \citet{Cuzzietal2003}. All these components are set in a \textit{matrix} of micron-sized grains. Radiometric dating suggests that CAIs are the oldest solar system solids, with an age of $\sim$4568 Myr, while chondrule formation occured 1-3 Myr later \citep[e.g.][]{BouvierWadhwa2010,Villeneuveetal2009,Amelinetal2010}.

  While the abundances of nonvolatile elements 
 in all chondrites roughly match that of the solar photosphere \citep{PalmeJones2005}, chondrites do present a diversity in their chemistry, the isotopic composition of several elements, most prominently oxygen \citep{Yurimotoetal2008}, and in the textures and abundances of their different components 
 \citep{BrearleyJones1998}. 
14 discrete chondrite \textit{groups} have been hitherto recognized, each of which representing either a single parent body or an array of parent bodies from the same reservoir.

 Meteoriticists have basically envisioned two possible explanations for this diversity
: The first one is that the reservoirs represented by each chondrite group acquired different compositions for their condensed fraction as a result of differing thermal histories
. Little mixing is assumed between these reservoirs. 
Since differences between chondrite groups would be largely traced to what did and did not condense out of the initially hot gas, this ``regional'' view has come to be known as the \textit{incomplete condensation model} \citep[e.g.][]{WassonChou1974,Blandetal2005}. The second possibility is that mixing was extensive in the solar nebula, and that chondrite diversity results from the incorporation of essentially the \textit{same} components, but in different \textit{proportions} depending on the group. Then, the different components need not be cogenetic. The latter would schematically comprise a high-temperature component (chondrules and CAIs) and a low-temperature component (matrix). 
 This is the \textit{two-component model} \citep[e.g.][]{Anders1964,Zandaetal2006}.

  Along with the mechanism(s) that caused \textit{fractionation}---i.e. separation of compositionally distinct constituents---in chondrites, it remains to be understood whether their diversity primarily reflects variations in space (i.e., heliocentric distance) or time. \citet{Rubin2011} illustrates the former view, with chondrites roughly arranged in order of increasing oxidation state with increasing heliocentric distance, ending with carbonaceous chondrites. \citet{Cuzzietal2003} and \citet{Chambers2006} advocate the latter view, with carbonaceous chondrites being the oldest. 

   The age range spanned by the different inclusions within single chondrites suggests that they spent up to several Myr as free-floating objects in the accretion disk. 
 Significant fractionation may thus have resulted from their transport dynamics between the times of their formation and their accretion. In most disk models, these dynamics are governed by gas-grain interaction \citep{CuzziWeidenschilling2006}. This is supported by petrographic evidence: Chondrules and other non-matrix chondrite components in a given meteorite exhibit a narrow distribution as a function of the product of the density and the radius, which is the relevant aerodynamic parameter \citep{CuzziWeidenschilling2006,KueblerMcSween1999,Mayetal1999}. Aerodynamic sorting could also account for metal-silicate fractionation \citep[see e.g.][]{Zandaetal2006}. 
Finally, the small size ($\lesssim 20\:\rm \mu m$) of high-temperature minerals identified in cometary material 
 is consistent with the increasing difficulty of outward aerodynamic transport of inner solar system material with increasing size \citep{HughesArmitage2010}. 

  Several studies have addressed the dynamics of solids within the disk \citep{CuzziWeidenschilling2006}, with focus on refractory inclusions or other high-temperature materials (e.g. \citealt{
BockeleeMorvanetal2002, WehrstedtGail2002,Cuzzietal2003,
Ciesla2009b,Ciesla2009,Ciesla2010,HughesArmitage2010}), the incomplete condensation model \citep{Cassen1996,Cassen2001,Ciesla2008}, water transport \citep[e.g.][]{CieslaCuzzi2006}. 

  In this paper we revisit the radial dynamics of chondrite components from a simplified, analytical point of view, taking the above works as guides. Our goal is to find out whether the diversity of chondrites can be understood in terms of the transport of their components in the disk, adopting standard modeling assumptions. We will show that inferences can be made in this respect independently of the details of the structure or turbulent properties of the disk, which are still uncertain.

  The outline of this paper is as follows: In Section \ref{Super-clans}, we briefly review a first-order classification of chondrites in carbonaceous and non-carbonaceous chondrites. We review the dynamics of gas and solids in Sections \ref{Disk modeling} and \ref{Solids}, respectively. In Section \ref{Net flux}, we introduce the ``gas-solid decoupling parameter'' $S$ central to this study and justify an approximation of the one-dimensional, vertically integrated continuity equation. 
In Section \ref{Arguments}, we present four arguments supporting the conjecture that non-matrix chondrite components had $S<1$ when carbonaceous chondrites accreted and $S>1$ when non-carbonaceous chondrites accreted. Cosmochemical implications are discussed in Section \ref{Implications}. In Section \ref{Conclusion}, we conclude.

\section{A first-order classification of chondrites}
\label{Super-clans}

\begin{figure}
\resizebox{\hsize}{!}{
\includegraphics{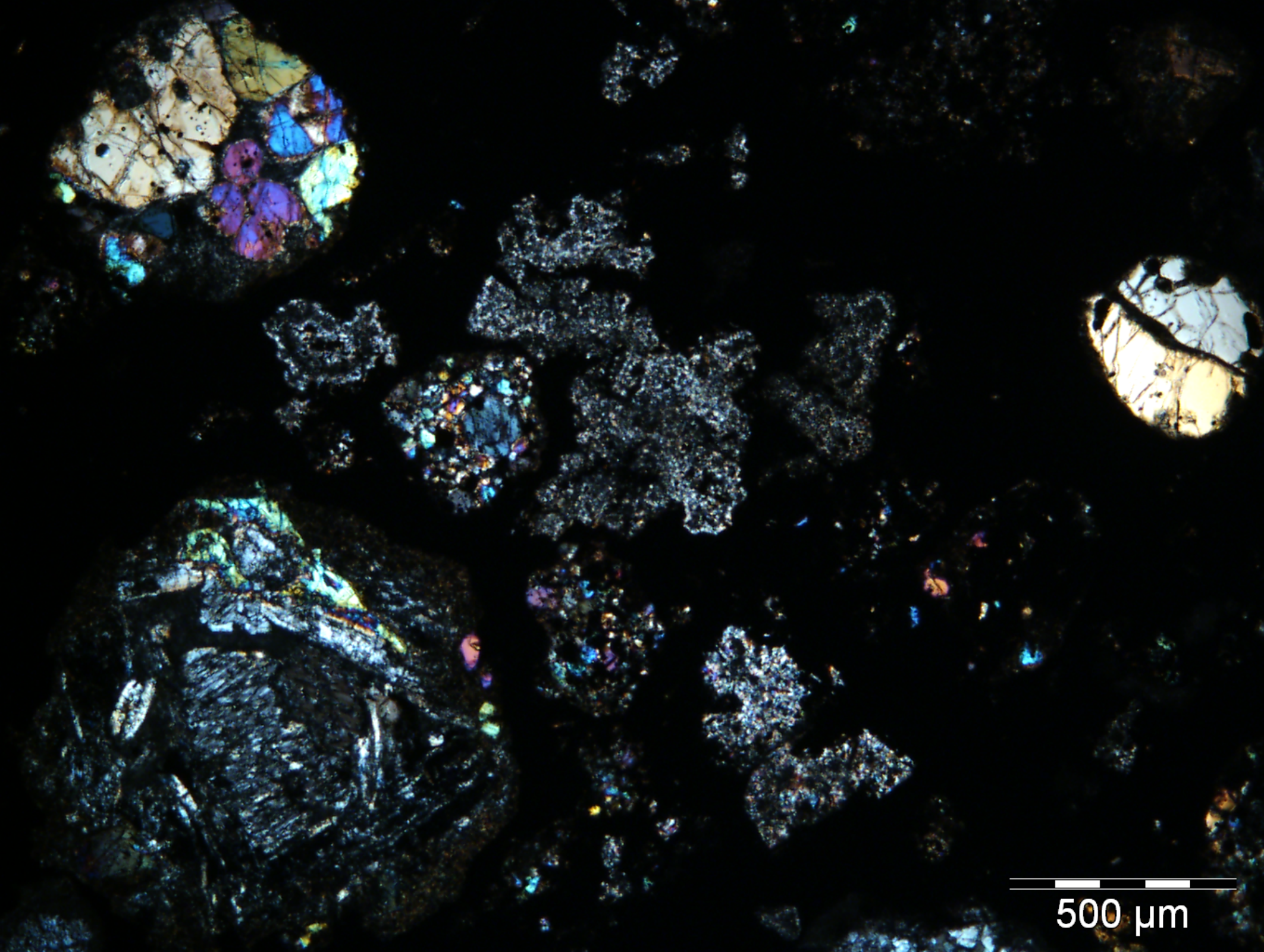}
}
\resizebox{\hsize}{!}{
\includegraphics{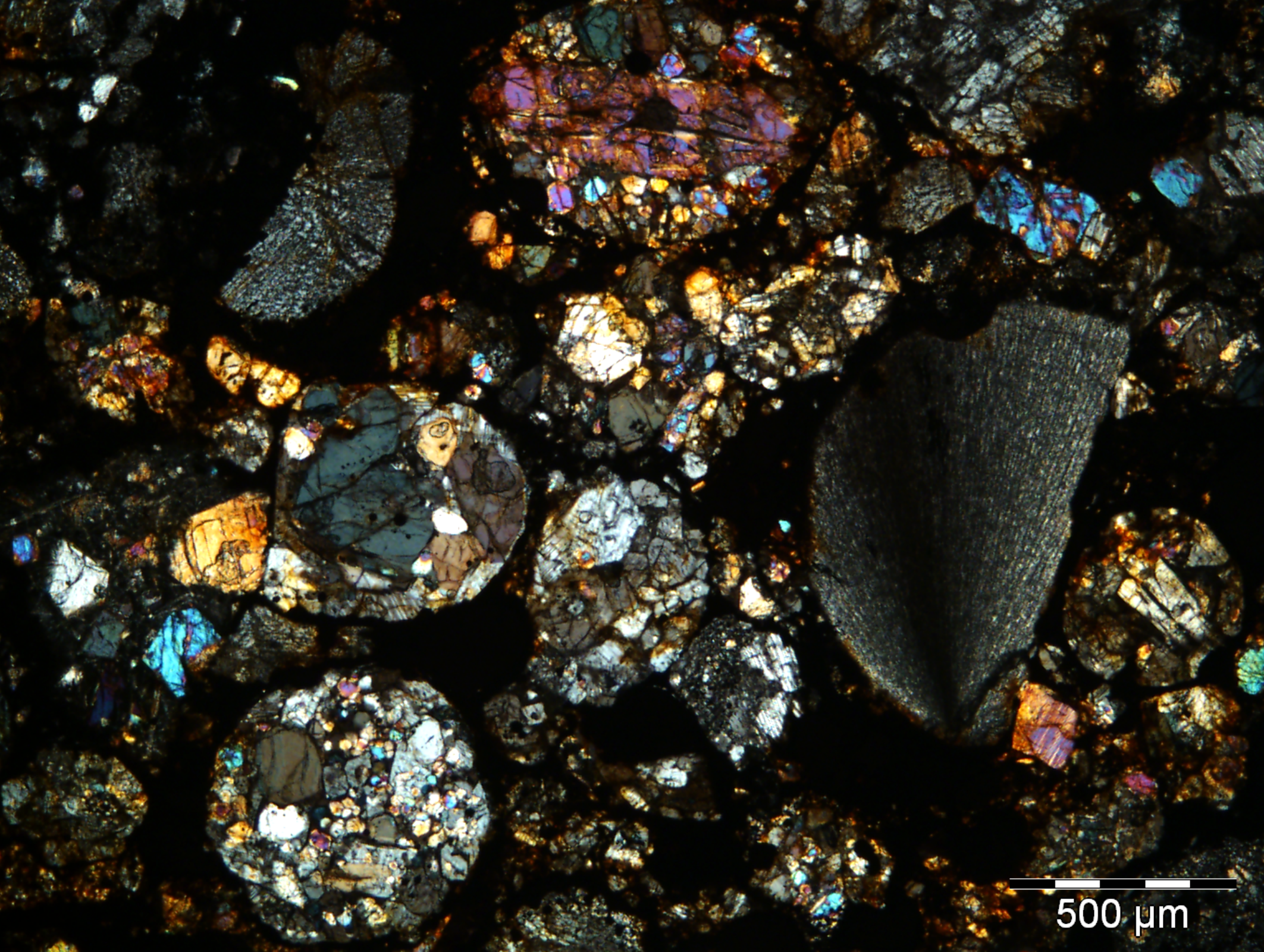}
}
\caption{Photomicrographs of thin sections of a carbonaceous chondrite (Allende, top) and an ordinary chondrite chondrite (Hallingeberg, bottom), viewed in polarized light. The round objects (sometimes fragmented), generally with colorful crystals, are chondrules. Irregularly-shaped, fine-grained objects with bluish-whitish shades visible in the upper photograph (mostly near the center) are CAIs. All these objects are set in a matrix (black in transmitted light).}
\label{photos}
\end{figure}

\begin{table}
\caption{Compared primary properties of chondrites}
\label{CC vs EOR}
\begin{tabular}{c c c}
\hline \hline
Properties & C chondrites & EOR chondrites\\
\hline
CAIs & abundant & rare\\
matrix fraction & high & low\\
monotonic trend of lithophile & &\\ elements with volatility? & yes & no\\
Mg/Si & solar & subsolar\\
chondrule O isotopes  & $^{16}$O-rich & $^{16}$O-poor\\
--- & variable & less variable\\
\hline
\end{tabular}
\end{table}

  It is certainly beyond our scope to propose particular geneses for each chondrite group. Instead, we shall content ourselves with a first-order classification of chondrites in two \textit{super-clans} \citep{Kallemeynetal1996,Warren2011b}: 
 the carbonaceous chondrites and the non-carbonaceous chondrites (see Figure \ref{photos} and Table \ref{CC vs EOR}).  

 The \textit{carbonaceous chondrites} (hereafter the ``CCs'') comprise the CI, CM, CO, CV, CK and CR groups
---we leave aside the CH and CB chondrites, which may have formed in very peculiar environments 
(e.g. \citet{Krotetal2005} but see \citealt{Gounelleetal2007})---and are the most chemically primitive chondrites. The \textit{non-carbonaceous chondrites} comprise the \textit{E}nstatite (EH, EL), the \textit{O}rdinary (H, L, LL) and the \textit{R}umuruti-type (R) chondrites and will be henceforth be referred to as ``EORs''
. Schematically, the main distinguishing features are (see \citet{BrearleyJones1998,ScottKrot2003} for detailed descriptions of chondrites): 

(i) CCs have higher CAI abundances (up to $>5$ \% in volume; e.g. \citealt{Hezeletal2008}) than EORs ($<$0.1 \%).
 
(ii) CCs have a higher proportion of (fine-grained) matrix (generally $>$30 \%) 
 than EORs ($\gtrsim 10$ \%). Chondrule textures also testify to higher 
 dust concentrations in their formation environment for CCs \citep{Rubin2010}. 

(iii) CCs exhibit a depletion (compared to solar abundances) of moderately volatile elements relative to refractory elements, as an essentially monotonic function of the 50 \% condensation temperature, while EORs exhibit no such clear correlation (except for the most volatile elements).
 
(iv) The Mg/Si ratio is solar for CCs and subsolar for EORs. 
 
(v) CCs have more $^{16}$O-rich chondrules than EORs. Also, in the three-isotope diagram, chondrules from single CCs define larger fields, which partly overlap among different chemical groups, while chondrules from single EORs form narrower, resolved fields \citep{Rubin2000, Clayton2003, Jonesetal2000}.

 Regarding CCs, while properties (iii) and (iv) are true of them as whole rocks, they do not seem to hold for their individual components 
 separately (\citet{Blandetal2005,HezelPalme2010}, but see \citealt{Zandaetal2012}). This is known as \textit{complementarity}.

  A primary assumption in this work will be that that the proportions of components in chondrites reflect their relative abundances around the disk midplane prior to accretion. 
It is however conceivable \citep[e.g.][]{Ciesla2010} that 
 the accretion process (if not gravity-controlled) 
 caused preferential incorporation of components 
  of certain sizes: This we shall refer to as the ``accretion bias''. A prototypical example is the turbulent concentration scenario of \citet{Cuzzietal2001}, but similar sorting calculations would be desirable e.g. for the streaming instability or for ``hit-and-stick'' 
 growth. 
 The evidence for complementarity in CCs 
 would militate against any accretion bias between chondrules and matrix grains. At any rate, the lack of a systematic chondrule size difference between CCs and EORs suggests that 
accretion bias would not account for the \textit{differences} between the two super-clans as wholes. 

\section{Disk modeling}
\label{Disk modeling}

The disk is described in a cylindrical coordinate system, with $R$ the heliocentric distance, $z$ the altitude above the midplane, and $\phi$ the azimuthal angle. We denote by $\mathbf{u}$, $\rho$, $T$ and $P=\rho c_s^2$ the gas velocity, density, temperature and pressure, respectively, with $c_s=\sqrt{k_BT/m}$ the isothermal sound speed, where $k_B$ is the Boltzmann constant and $m$ the mean molecular mass (here 2.33 times the proton mass). $\Omega$ and $v_K\equiv\Omega R$ are the Keplerian angular and linear velocities, respectively.
 
  We treat the disk as vertically isothermal (see Appendix A), and thus vertical hydrostatic equilibrium implies:
\begin{equation}
\rho(R,z)=\frac{\Sigma(R)}{\sqrt{2\pi}H(R)}\exp{\left(-\frac{z^2}{2H(R)^2}\right)}
\label{density stratification}
\end{equation}
with the pressure scale height $H=c_s/\Omega$ and  the surface density $\Sigma\equiv\int_{-\infty}^{+\infty}\rho\mathrm{d}z$.

 We assume the disk to be turbulent 
and assume axisymmetry in the sense that variations of any quantity $Q$ in the azimuthal direction may be treated as turbulent fluctuations about a mean $\overline{Q}$. We denote the Eulerian perturbations with $\delta Q\equiv Q-\overline{Q}$. 


  In that case, from angular momentum conservation
:
\begin{equation}
\overline{u_R}=-\frac{2}{\rho}\left[\frac{1}{R^{1/2}}\frac{\partial}{\partial R}\left(\frac{R^{1/2}}{\Omega}T_{R\phi}\right)+\frac{1}{\Omega}\frac{\partial T_{z\phi}}{\partial z}\right]\sim -\alpha \frac{c_s^2}{v_K}
\label{uR}
\end{equation}
with $T_{R\phi}$ and $T_{z\phi}$ the $R\phi$ and $z\phi$ components of the turbulent stress tensor \citep{BalbusPapaloizou1999}\footnote{Noting $\mathbf{B}$ the magnetic field, they are defined as ($i=R,z$):
\begin{eqnarray}
T_{i\phi}\equiv \rho\overline{\delta u_\phi\delta u_i}-\frac{\overline{B_\phi B_i}}{\mu_0}.\nonumber
\end{eqnarray}}
and $\alpha$ a dimensionless number parameterizing the former as:
\begin{equation}
T_{R\phi}=\frac{3}{2}\alpha P
\end{equation}
It must be noted that defined as such, $\alpha$ may \textit{a priori} vary in space and time, and in particular, we do not make the assumption that the stress tensor obeys a viscous prescription \citep[e.g.][]{TakeuchiLin2002}. Observations of disks suggest $\alpha$ values around $10^{-2}$ at 10-100 AU scales \citep{Hartmannetal1998}, in the range of those achieved by the magneto-rotational instability \citep{BalbusHawley1998,LesurLongaretti2007,BaiStone2011}. In the dead zone \citep{Gammie1996}, where the latter is unlikely to operate, $\alpha$ should drop to $10^{-6}-10^{-4}$ around the midplane \citep[e.g.][]{FlemingStone2003,IlgnerNelson2008,OishiMcLow2009,Turneretal2010}.

  The mass accretion rate can then be expressed as:
\begin{equation}
\dot{M}\equiv-2\pi R\int_{-\infty}^{+\infty}\rho\overline{u_R}\mathrm{d}z =6\pi R^{1/2}\frac{\partial}{\partial R}\left(R^{1/2}\Sigma\nu\right)
\label{mass accretion rate}
\end{equation}
where 
\begin{equation}
\nu\equiv\frac{1}{\Sigma}\int_{-\infty}^{+\infty}\alpha \frac{c_s^2}{\Omega}\rho\mathrm{d}z\equiv\langle\alpha\rangle_\rho\frac{c_s^2}{\Omega}, 
\end{equation}
with $\langle\alpha\rangle_\rho$ the density-weighted vertical average of $\alpha$.

Note that we hereby ignore infall (unlike e.g. \citealt{Zhuetal2010b}) and photoevaporation (unlike e.g. \citealt{Desch2007,HughesArmitage2010}), which may be most important in the early and late stages of the disk, respectively. However, while these effects would influence the evolution of $\Sigma$, 
they would not affect the validity of the radial velocity formula (\ref{uR}) in the disk interior.

An important benchmark for this paper is the steady-state disk, where $\dot{M}$ is uniform. This  
should be a good approximation of 
 the inner regions of a real disk if the local evolution 
 timescale
\begin{equation}
t_{\rm vis}(R)\equiv\frac{R^2}{\nu}=0.2\:\mathrm{Myr}\:R_{\rm AU}^{1/2}\left(\frac{1\:\mathrm{km/s}}{c_s}\right)^2\left(\frac{10^{-3}}{\langle\alpha\rangle_\rho}\right)
\label{tvis}
\end{equation}
(which is the timescale of accretion to the Sun of gas at heliocentric distance $R$) is shorter than the disk's age. 
In that case, if neither $\Sigma$ nor $\nu$ diverges as $R\rightarrow 0$, we have: 
\begin{equation}
\dot{M}=3\pi\Sigma\nu
\label{steady mass accretion rate}
\end{equation}

\section{Dynamics of solid particles}
\label{Solids}
We now proceed to the transport of solid particles. By ``particle'', we mean any detached solid body, belonging to any of the chondrite components mentioned in Section \ref{Introduction}. We first review the dynamics for a single particle, before writing the mean-field continuity equation for a population of identical particles.
\subsection{Single particle dynamics}
The dynamics of a small particle are dominated by gas drag. For a spherical solid particle of radius $a$ and internal density $\rho_s$, the stopping time is \citep{Epstein1924} 
\begin{equation}
\tau=\sqrt{\frac{\pi}{8}}\frac{\rho_sa}{\rho c_s}.
\label{Epstein}
\end{equation}
For a 0.3 mm-radius chondrule, $\rho_sa=1\:\mathrm{kg/m^2}$, which we take as our normalizing value. Within a factor of a few, it indeed pertains to most chondrules, metal/sulfide grains and refractory inclusions alike in most chondrite groups \citep{KingKing1978,BrearleyJones1998,KueblerMcSween1999,Mayetal1999}. 
 We do not study here the fate of larger ($\gtrsim$ 1 cm) bodies. This is because (unbrecciated) chondrites are homogeneous at the centimeter scale and thus their composition was determined at the agglomeration of their millimeter-sized components\footnote{However, some of these larger bodies may have (i) travelled to another reservoir and (ii) contributed by shattering to its sub-millimeter-size 
 inventory. This is thus ignored here.}. 

  A measure of the coupling of the particles to the gas on an orbital timescale is the dimensionless stopping time
\begin{equation}
\textrm{St}\equiv\Omega\tau=\frac{\pi}{2}\frac{\rho_sa}{\Sigma}=2\times 10^{-4}\left(\frac{\rho_sa}{1\:\mathrm{kg/m^2}}\right)\left(\frac{10^3\:\mathrm{g/cm^2}}{\Sigma}\right),
\label{St}
\end{equation}
where the second equality holds at the disk midplane. Clearly, $\mathrm{St}\ll 1$ for millimeter- or centimeter-sized particles and smaller.  
In this limit, gas-solid drag forces the particles to essentially follow the gas, but with a systematic relative drift due to the pressure gradient, so that the particle velocity is given by 
 \citep{YoudinGoodman2005}: 
\begin{equation}
\mathbf{v}_p=\mathbf{u}+\tau\frac{\nabla P}{\rho},
\label{TVA}
\end{equation}
The radial component of the second term on the right-hand side,
\begin{equation}
 v_{\rm drift}\equiv\frac{\tau}{\rho}\frac{\partial P}{\partial R}\sim -\textrm{St}\frac{c_s^2}{v_K},
\label{vdrift}
\end{equation}
 is robustly negative in disk models \citep[except perhaps around the inner edge of the dead zone, e.g.][]{Dzyurkevichetal2010}, and alone entails a sunward drift on a timescale
\begin{equation}
t_{\rm drag}=\frac{R}{|v_{\rm drift}|}\sim\frac{1}{\Omega \textrm{St}}\left(\frac{v_K}{c_s}\right)^2=2\:\mathrm{Myr}\:R_{\rm AU}^{1/2}\left(\frac{1\:\mathrm{km/s}}{c_s}\right)^2\left(\frac{10^{-4}}{\textrm{St}}\right).
\label{tdrag}
\end{equation}

\subsection{Averaged continuity equation for a fluid of particles}
\label{turbulence averaged}
We now consider a population of identical, non-interacting particles and treat it as a fluid of density $\rho_p$ (not to be confused with the \textit{internal} density $\rho_s$ of each particle). 
The continuity equation averaged over turbulent fluctuations may be written in the form \citep{SchraeplerHenning2004,Balbus2009}:
\begin{eqnarray}
\frac{\partial\overline{\rho_p}}{\partial t}+\frac{1}{R}\frac{\partial}{\partial R}\left[R\left(\overline{\rho_p}\:\overline{v_{\rm p, R}}-D_{RR}\rho\frac{\partial}{\partial R}\overline{\left(\frac{\rho_p}{\rho}\right)}\right)\right]\nonumber\\+\frac{\partial}{\partial z}\left[\overline{\rho_p}\:\overline{v_{\rm p, z}}-D_{zz}\rho\frac{\partial}{\partial z}\overline{\left(\frac{\rho_p}{\rho}\right)}\right]=0.
\label{continuity 2D}
\end{eqnarray}
We shall henceforth drop the overbars. For $\textrm{St}\ll 1$, the diffusion coefficients equal those of passively advected scalars in the gas \citep{YoudinLithwick2007} and will be parameterized as:
\begin{eqnarray}
D_{RR}=\delta_R\frac{c_s^2}{\Omega}\:\:\:\:\:\:\mathrm{and}\:\:\:\:\:\:
D_{zz}=\delta_z\frac{c_s^2}{\Omega}.
\end{eqnarray}
For correlation times of order $\Omega^{-1}$ \citep{FromangPapaloizou2006,CuzziWeidenschilling2006}, $\delta_R$ and $\delta_z$ should be comparable to $\alpha$ because of the coupling between azimuthal and radial components due to Coriolis forces. 
We coin the radial and vertical Schmidt numbers\footnote{There is some variation in terminology in the litterature: This number is called a ``Prandtl number'' by \citet{Prinn1990} and \citet{CuzziWeidenschilling2006}, and  \citet{YoudinLithwick2007} use ``Schmidt number'' for the ratio of the diffusivity of gas to that of the particles (1 for $\textrm{St}\ll 1$).}:
\begin{eqnarray}
\textrm{Sc}_R\equiv\frac{\alpha}{\delta_R}\:\:\:\:\:\:\mathrm{and}\:\:\:\:\:\:
\textrm{Sc}_z\equiv\frac{\alpha}{\delta_z}.
\end{eqnarray}
\citet{Prinn1990} and \citet{DubrulleFrisch1991} argued that, in hydrodynamical turbulence, $\textrm{Sc}_R$ should not exceed 1, as is the 0.176 value obtained in turbulent rotating flow experiments by \citet{Lathropetal1992} or the $\sim$0.7 values mentioned by \citet{Gail2001}
. \citet{Johansenetal2006} reported $\textrm{Sc}_R=4.6\alpha^{0.26}$ and $\textrm{Sc}_z=25.3\alpha^{0.46}$ from local MHD net-flux simulations, which thus can exceed 1
, consistent with most earlier results \citep{Johansenetal2006}. These uncertainties make it prudent to keep track of the Schmidt numbers throughout this study, although we generally expect them to be of order unity.

\section{Net radial transport of solid particles}
\label{Net flux}
We now wish to derive the vertically-integrated continuity equation of the particle fluid. Beforehand, we will introduce the dimensionless numbers of special relevance in this problem and briefly review the vertical distribution of the particles.
\subsection{The gas-solid decoupling parameter $S$ }
\label{S number}
  The ratio of particle-gas drift and gas advection contributions in the radial transport is, using equations (\ref{uR}) and (\ref{vdrift}):
\begin{eqnarray}
\frac{v_{\rm drift}}{u_R} \:  \sim  \: \frac{\textrm{St}}{\alpha} \: \equiv \: S & = & \frac{\pi}{2}\frac{\rho_sa}{\Sigma\alpha}\\
 & = & 0.2\left(\frac{\rho_sa}{1\:\mathrm{kg/m^2}}\right)\left(\frac{10^{-3}}{\alpha}\right)\left(\frac{10^3\:\mathrm{g/cm^2}}{\Sigma}\right),\nonumber
\label{drift vs advection}
\end{eqnarray}
the last two equalities holding at the midplane. We also define:
\begin{eqnarray}
S_R\equiv\frac{\textrm{St}}{\delta_R}=S\times \textrm{Sc}_R\:\:\:\:\:\:\mathrm{and}\:\:\:\:\:\:\:
S_z\equiv\frac{\textrm{St}}{\delta_z}=S\times \textrm{Sc}_z
\end{eqnarray}
As we shall see in the next subsection, $S_z$ is a measure of vertical settling \citep{Cuzzietal1996}\footnote{This is what \citet{Cuzzietal1996} called $S$ (they had taken $\textrm{Sc}_z=1$).}, and we will see later in Section \ref{Argument 1} that $S_R$ characterizes the ratio of the particle-gas drift and the diffusion contributions to the radial flux. 

  So long the Schmidt numbers are of order unity, $S$, $S_R$ and $S_z$ actually are of the same order of magnitude. Hence, the qualitative aerodynamics of the particles are essentially controlled by one single number, namely $S$. Schematically, finite-size effects can be ignored for $S\ll 1$, while particles significantly decouple from the gas for $S\gg 1$. In that sense, one may call $S$ the ``gas-solid decoupling parameter''. 
 To keep track of the exact role of $S$, $S_R$ and $S_z$, we shall however remain careful in writing either one in the formulas (unless otherwise noted, they will be implicitly understood as being evaluated at the midplane).
\subsection{Vertical distribution of particles}
\label{vertical distribution}
The vertical flux provides the dominating balance in equation (\ref{continuity 2D}). On timescales longer than the vertical mixing timescale 
\begin{equation}
t_{\rm vm}=\frac{1}{\Omega\:\mathrm{max}(\delta_z,\textrm{St})},
\end{equation}
the vertical distribution of the particles obeys:
\begin{equation}
\frac{\partial}{\partial z}\mathrm{ln}\frac{\rho_p}{\rho}=-S_z(R,z)\frac{z}{H^2},
\label{vertical equilibrium}
\end{equation}
from which the role of $S_z$ as a measure of settling is obvious, with dimensional analysis indicating a dust layer thickness of order $H/\sqrt{1+S_z}$ \citep[e.g.][]{Dubrulleetal1995}.

\subsection{One-dimensional continuity equation}
We now wish to integrate the two-dimensional continuity equation (\ref{continuity 2D}) over $z$ to obtain the evolution equation of the particle surface density $\Sigma_p\equiv\int_{-\infty}^{+\infty}\rho_p\mathrm{d}z$. We must thus calculate the $\rho_p$-weighted average of the particle velocity, similarly to \citet{TakeuchiLin2002}. For any quantity $Q$ and any weight function $w(z)$, we define its $w$-weighted average as:
\begin{equation}
\langle Q\rangle_w\equiv\frac{\int_{-\infty}^{+\infty}Q(z)w(z)\mathrm{d}z}{\int_{-\infty}^{+\infty}w(z)\mathrm{d}z}.
\end{equation}
  $v_{\rm drift}(z)$ can be readily expressed from equations (\ref{vdrift}) and (\ref{density stratification}).
Same does not hold true, however, for $u_R$. As it stands, equation (\ref{uR}) expresses it in terms of turbulent correlations whose dependence on $z$ 
are essentially unknown \citep[e.g.][]{MillerStone2000,Fromangetal2011}. Moreover, 
the nature of the turbulence may change with height \citep[e.g.][]{FlemingStone2003,Turneretal2010}. In the current state of accretion disk theory, the vertical flow structure is thus uncertain.

  However, as we shall show, the net flow of the particles is actually \textit{weakly sensitive} to the vertical gas flow structure, provided $\textrm{Sc}_R\sim\textrm{Sc}_z\sim 1$. In the limit $S_z\gg 1$, the particles concentrate around the midplane so that
\begin{equation}
\langle v_{\rm p,R}\rangle_{\rho_p}\approx u_R(R,0)+v_{\rm drift}(R,0)\approx v_{\rm drift}(R,0),
\label{approximation settled}
\end{equation}
where the latter approximation uses the scaling $v_{\rm drift}/u_R\sim S$ (and hereby the hypothesis that $\textrm{Sc}_z\sim 1$)
. In the limit $S_z\ll 1$, particles are well mixed vertically, and we have
\begin{equation}
\langle v_{\rm p,R}\rangle_{\rho_p}\approx 
\langle u_R\rangle_{\rho_p}
\approx \langle u_R\rangle_\rho=-\frac{3}{\Sigma R^{1/2}}\frac{\partial}{\partial R}\left(R^{1/2}\Sigma\nu\right).
\label{approximation well-mixed}
\end{equation}
 We thus see that the approximation 
\begin{equation}
\langle v_{\rm p,R}\rangle_{\rho_p}\approx \langle u_R \rangle_\rho +v_{\rm drift}(R,0)
\label{approximation}
\end{equation}
should be accurate in either limit. For $S\sim 1$, the approximations used break down, but, unless $\textrm{Sc}_z$ is very different of 1, or $\langle\alpha\rangle_\rho$ greatly exceeds $\alpha$ near the midplane
, the imparted error should not exceed a factor of a few then.

  As regards the diffusion term, vertical integration yields:
\begin{eqnarray}
-\int_{-\infty}^{+\infty}D_{RR}\rho\frac{\partial}{\partial R}\left(\frac{\rho_p}{\rho}\right)\mathrm{d}z=-\langle D_{RR}\rangle_{\rho_p}\Sigma\frac{\partial}{\partial R}\left(\frac{\Sigma_p}{\Sigma}\right)\nonumber\\ - \Sigma_p\int_{-\infty}^{+\infty}D_{RR}\frac{\partial f}{\partial R}\frac{\rho\mathrm{d}z}{\Sigma}
\label{integrate diffusion}
\end{eqnarray}
where $f$ is defined by
\begin{equation}
\frac{\rho_p(R,z)}{\rho (R,z)}\equiv\frac{\Sigma_p (R)}{\Sigma (R)}f(R,z).
\label{f defined}
\end{equation}
In principle, the second term on the right-hand-side of equation (\ref{integrate diffusion}) should be included as an effective correction to $\langle v_{\rm p,R}\rangle_{\rho_p}$ in the net flux. 

However, this correction is negligible compared to $v_{\rm drift}(R,0)$ if $S\gg 1$ (as it is then of order $\langle u_R\rangle_\rho$) and negligible compared to $\langle u_R\rangle_\rho$ if $S\ll 1$ (as $f\approx 1$ so that $\partial f/\partial R\ll 1$). 
We shall thus drop it from the vertically-integrated flux.

  Hence, the one-dimensional continuity equation sought is:
\begin{eqnarray}
\frac{\partial\Sigma_p}{\partial t}+\frac{1}{R}\frac{\partial}{\partial R}\bigg(R\bigg(\Sigma_p(\langle u_R \rangle_\rho+v_{\rm drift}(R,0))\nonumber\\-\langle D_{RR}\rangle_{\rho_p}\Sigma\frac{\partial}{\partial R}\left(\frac{\Sigma_p}{\Sigma}\right)\bigg)\bigg)=0
\label{Sigmap evolution}
\end{eqnarray}
We will henceforth drop the $\langle ...\rangle_{\rho,\rho_p}$ and refer to $v_{\rm drift}(R,0)$ simply as $v_{\rm drift}$. This equation is essentially that used in one-dimensional models in the literature \citep[e.g.][]{Cuzzietal2003, Ciesla2010, HughesArmitage2010}
. We have shown here that it is accurate (within order-unity corrections, up to a factor of a few for $S\sim 1$), quite insensitively to the as yet uncertain details of turbulence modeling, and thus most appropriate to evaluate net radial transport. 

\section{The value of $S$ at chondrite accretion}
\label{Arguments}


  In this section we shall advocate the following conjecture:
\\
\\ \textit{Non-matrix chondrite components had $S<1$ at the accretion of CCs and $S>1$ at the accretion of EORs.}
\\
\\  Since $S$ depends on size, we stress that we are speaking here of the largest, compositionally distinctive chondritic components, that is chondrules, CAIs, metal and sulfide grains
. As mentioned in the introduction, the range of $\rho_sa$ of these inclusions in single chondrites is narrow, and roughly independent of the nature of the inclusions \citep[e.g.][]{KingKing1978,KueblerMcSween1999,Mayetal1999}, such that speaking of \textit{one} value of $S$ for a given chondrite should not be ambiguous. Moreover, mean chondrule sizes of different chondrite groups do not vary by more than a factor of a few \citep{BrearleyJones1998} 
 so that, given the order-of-magnitude variations 
 of $\Sigma$ and $\alpha$, we can even schematically think of \textit{all} these inclusions in \textit{all} chondrites as having the same $\rho_sa\approx 1\:\rm kg/m^2$. Therefore, variation in the value of $S$ at the accretion of the different chondrites will \textit{essentially} reflect differences in gas properties  rather than intrinsic variations of the acccreted components. 
 
 The above conjecture is supported by four arguments which we shall now examine each in turn. In each subsection, we begin with a property of $S$ (all listed in Table \ref{S properties}) and then link it to one of the meteoritic observations of Section \ref{Super-clans}. 

\begin{table}
\caption{Properties of $S$}
\label{S properties}
\begin{tabular}{c c c c}
\hline \hline
Property & $S < 1$ & $S > 1$ & Section\\
\hline
Significant outward transport? & yes & no & \ref{Argument 1}\\
Settling of mm-sized bodies? & no & yes & \ref{Argument 2}\\
Coherence between grains? & yes & difficult & \ref{Argument 3}\\
Radial diffusion & unlimited & limited & \ref{Argument 4}\\
Proximity to Sun/epoch & close/early & far/late & \ref{S steady disks}\\
\hline
\end{tabular}
\end{table}

\subsection{Outward transport and CAI abundance}
\label{Argument 1}

We wish to show here that particles formed close to the Sun cannot be transported much beyond the $S=1$ line in significant amounts, be it through outward advection or outward diffusion.

  As regards advection, the argument essentially relies on the scaling relationship (\ref{drift vs advection}) $v_{\rm drift}/u_R\sim S$ mentioned in Section \ref{S number}. For $S\gg 1$, the gas-drag induced drift velocity dominates and is robustly inward, and thus no significant outward transport is possible in this region, even if the net \textit{gas} flow is outward \citep[e.g.][]{Desch2007,Jacquetetal2011a}. Thus, if a given region contains a significant abundance of a given component with $S>1$, the latter cannot have been transported from the vicinity of the Sun to there but must have formed \textit{in situ} or further out, unless it was delivered by some mechanism other than the in-disk transport envisioned here (e.g. via jets as in \citealt{Shuetal2001}). 
One illustration are the ``midplane flow'' simulations 
 of \citet{HughesArmitage2010}
: for example, in their Fig. 5b2, 20 $\mu$m particles pile up at $\sim$90 AU, corresponding to $S=3.2$. 

  Turbulent diffusion would not alleviate this constraint. To offset particle-gas drift, it indeed requires the particle-to-gas ratio to vary on a lengthscale $R/S_R$, amounting to a steep inward gradient if $S_R\gg 1$. For example, \citet{HughesArmitage2012} found in their simulations that particles were confined to regions \textit{at most} with $\textrm{St}\leq 10^{-1}$, i.e. $S\leq 10$ for their $\alpha=10^{-2}$. 

   It is instructive to calculate the particle-to-gas ratio profile that results from balancing the different contributions of the net radial flux of the particles in equation (\ref{Sigmap evolution}), that is:
\begin{equation}
(u_R+v_{\rm drift})\Sigma_p-D_{RR}\Sigma\frac{\partial}{\partial R}\left(\frac{\Sigma_p}{\Sigma}\right)=0.
\end{equation}
From this, we obtain:
\begin{equation}
\frac{\partial}{\partial R}\mathrm{ln}\frac{\Sigma_p}{\Sigma}=\frac{u_R+v_{\rm drift}}{D_{RR}}=-3\textrm{Sc}_R\frac{\partial}{\partial R}\mathrm{ln}\left(R^{1/2}\Sigma\nu\right)+S_R\frac{\partial \mathrm{ln}P}{\partial R}.
\end{equation}
where we have injected equations (\ref{mass accretion rate}) and (\ref{vdrift}). Assuming $\textrm{Sc}_R$ to be constant
, this may be integrated to yield:
\begin{equation}
\frac{\Sigma_p}{\Sigma}\propto \frac{\exp{\left(\int^RS_R\frac{\partial\mathrm{ln}P}{\partial R}\mathrm{d}R'\right)}}{\left(\Sigma\nu R^{1/2}\right)^{3\textrm{\scriptsize Sc}_R}} 
\label{static CAI}
\end{equation} 
If we take the surface density of the \textit{gas} as a proxy for the total abundance of solids
, $\Sigma_p/\Sigma$ is proportional to the fraction of the component in question (say, CAIs) in chondrites at their accretion location.
 
\begin{figure}
\resizebox{\hsize}{!}{
\includegraphics{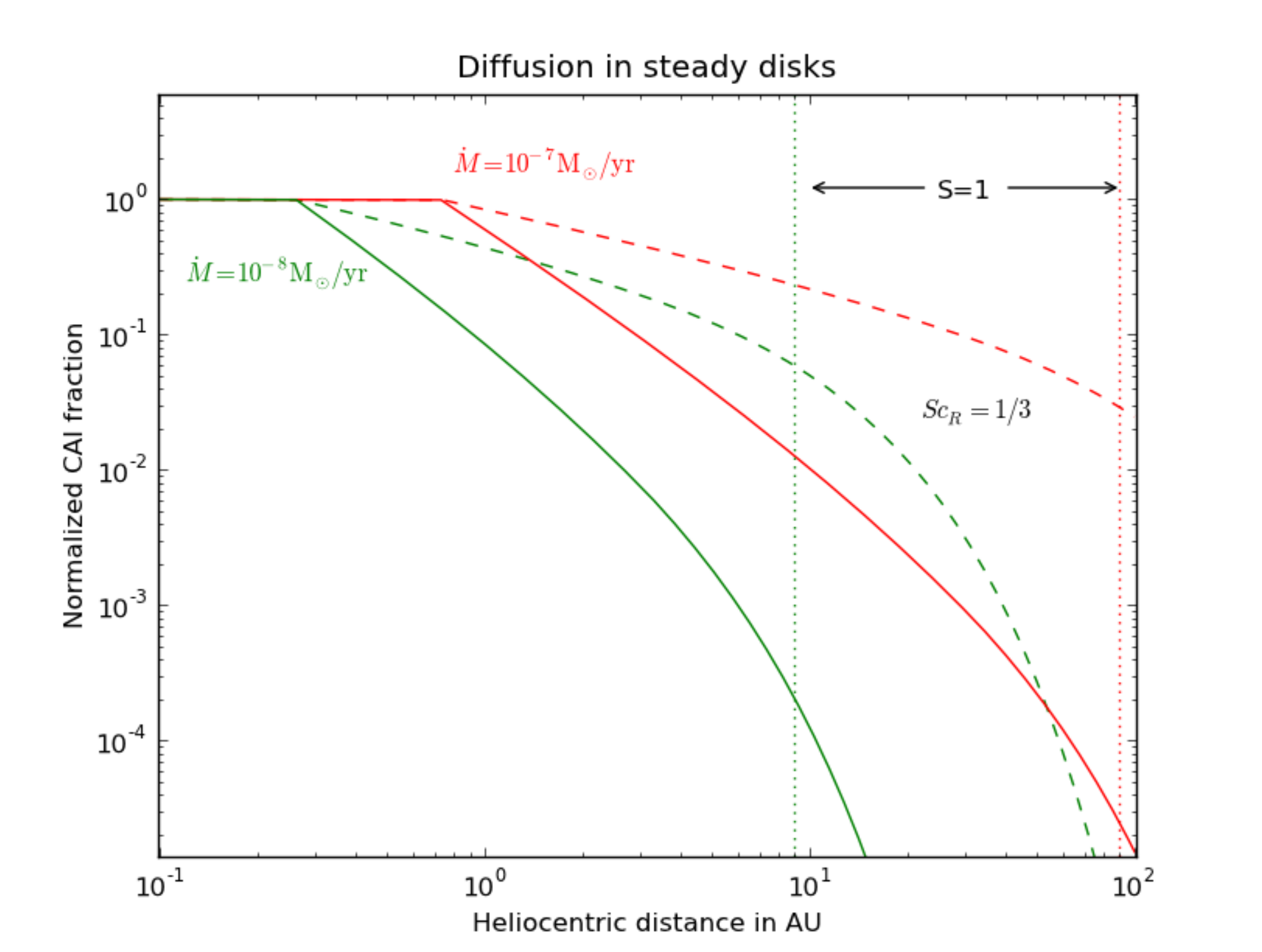}
}
\caption{Plot of the equilibrium diffusion profile of CAIs (or any chondrite component produced near the Sun) in a steady disk. The ``normalized CAI fraction'' is proportional to the CAI-to-gas ratio and is set to unity for $T>1500$ K. Results are shown for two mass accretion rates, $10^{-7}$ (red) and  $10^{-8}$ (green) $\mathrm{M_\odot . yr^{-1}}$. Solid lines assume $\textrm{Sc}_R=1$ and dashed ones $\textrm{Sc}_R=1/3$ (an enhanced radial diffusivity). Vertical dotted lines mark the heliocentric distance where $S=1$ for both values of $\dot{M}$. We have taken $f_T=1$, $\kappa = 5\:\rm cm^2/g$ and $\alpha = 10^{-3}$ (see Section \ref{S steady disks}).}
\label{diffusion}
\end{figure}

  To be more definite, we specialize to the case of a steady-state disk and assume radial dependences of the form $\Sigma\propto R^{-p}$ and $T\propto R^{-q}$
. Equation (\ref{static CAI}) becomes:
\begin{equation}
\frac{\Sigma_p}{\Sigma}\propto \left(\frac{\exp{(-cS)}}{R^{3/2}}\right)^{\textrm{\scriptsize Sc}_R},
\label{steady CAI}
\end{equation}
with $c\equiv (2p+q+3)/(3-2q)$ a number of order unity\footnote{We have used equation (\ref{S steady}). If $\alpha$ is constant (and thus $p+q=3/2$), $c=17/4$ and 11/4 for the viscous-heating- and the irradiation-dominated regimes, respectively (see Section \ref{S steady disks}).}. This is plotted in Fig. \ref{diffusion}
. For $S\ll 1$, the $R^{-3\textrm{\scriptsize Sc}_R/2}$ dependence matches that found by \citet{ClarkePringle1988} for a passive contaminant. The new result here is that $S_R=1$ clearly represents a cutoff for outward diffusion, but depending on $\textrm{Sc}_R$ (see Section \ref{turbulence averaged}), diffusion of inner solar system material may not necessarily have been efficient until even there, especially if the heliocentric distance must climb one or two orders of magnitude. 

  The $S=1$ line thus schematically represents the maximum range of outward transport within the disk. It is set by the disk model and may be located \textit{a priori}, before any simulation. 
Then, the high abundance of CAIs---which formed close to the Sun---in CCs relative to EORs implies $S<1$ for the former.


\subsection{Settling and proportion of matrix}
\label{Argument 2}
  We mentioned in Section \ref{vertical distribution} that $S_z$ was a measure of settling. While for $S_z\ll 1$, the particles may be considered to be well-mixed vertically, for $S_z \gg 1$, particles are concentrated around the midplane, with their particle-to-gas ratio being enhanced by a factor of order $\sqrt{S_z}$ \citep{Dubrulleetal1995}.

  Thus, the abundance of chondrules with $S_z\gg 1$ is enhanced relative to comparatively well-mixed micron-sized dust grains. The maximum enhancement is reached when the dust grains themselves start to settle, i.e., when $S_{z,\mathrm{dust}}=((\rho_sa)_{\rm dust}/(\rho_sa)_{\rm ch})S_z>1$, where the ``dust'' and ``ch'' subscripts refer to dust grains and chondrules, respectively: assuming similar densities for both, this maximum enhancement is $\sqrt{a_{\rm ch}/a_{\rm dust}}$, which, for micron-sized matrix grains and chondrule radii of a few tenths of millimeter (see Section 4.1), evaluates to $\sim$10. 
In view of this, the fact that EORs tend to have less matrix ($\lesssim 10$ \%) than CCs ($\gtrsim 30$ \%) would be accounted by $S_z\lesssim 1$ for CCs and $S_z\gg 1$ for EORs.

  Before closing this subsection, we comment that $S_z >1$ for EORs would also be consistent with a lower \textit{absolute} dust concentration in the formation region of their chondrules. Indeed, for a fixed $\delta_z$, a larger level of settling would correspond to a smaller column density of the gas, and hence of the solids for a given dust-to-gas column density ratio.\footnote{This would not be compensated by concentration at the midplane if $S_{\rm dust}> 1$ because then, defining $\epsilon_{\rm dust}\equiv\Sigma_{\rm dust}/\Sigma$, the dust density there is
\begin{eqnarray}
\Sigma_{\rm dust}\frac{\sqrt{S_{\rm z, dust}}}{\sqrt{2\pi}H}=\frac{\epsilon_{\rm dust}}{2H}\sqrt{\rho_sa_{\rm dust}\frac{\Sigma}{\delta_z}},\nonumber
\end{eqnarray}
which, everything else ($\epsilon_{\rm dust}$, $H$, $\delta_z$) being equal, \textit{decreases} with decreasing $\Sigma$.}

\subsection{Coherence of a grain population and complementarity}
\label{Argument 3}

\begin{figure}
\resizebox{\hsize}{!}{
\includegraphics{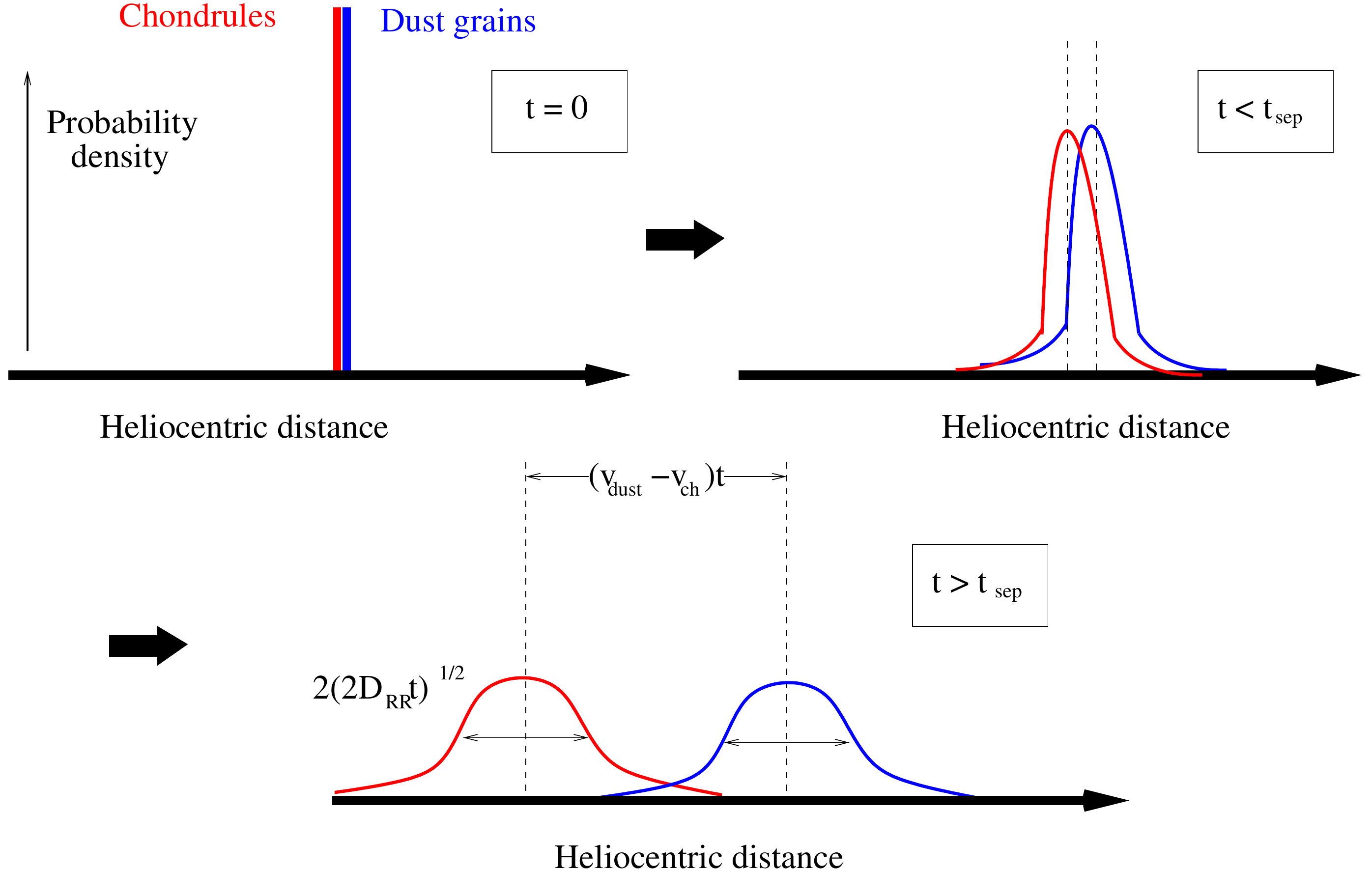}
}
\caption{Sketch of the decoherence of a chondrule population (red) and a dust grain population (blue) as a result of differential drift. We plot their probability density as a function of heliocentric distance after a chondrule formation event at time $t=0$ (when these probability densities are essentially delta functions). Due to turbulent diffusion, the two populations largely overlap for small $t$ despite a faster mean drift of chondrules (here velocities are denoted $v_{\rm ch}$ and $v_{\rm dust}$, respectively), but are spatially resolved after a timescale $t_{\rm sep}$. If accretion has not taken place by then, any chondrule-matrix complementarity will be lost.
}
\label{Complementarity}
\end{figure}

  Let us consider two particle populations, labelled 1 and 2, with different sizes, both concentrated at the same given heliocentric distance $R$ at $t=0$. This is sketched in Fig. \ref{Complementarity}. Although the differing sizes imply different mean velocities, these two populations do not instantaneously decouple from each other, because turbulent diffusion widens these distributions, such that they continue to overlap at short times. We call \textit{coherence} between two populations this state where both populations are not spatially resolved from one another. However, as the diffusional spread goes like $\sqrt{t}$ while the separation between the two peaks increases linearly with $t$, this coherence is expected to be lost by a timescale $t_{\rm sep}$ where the peaks are mutually resolved, and which we define with 
\begin{equation}
|v_{\rm drift, 1}-v_{\rm drift, 2}|t_{\rm sep}=2\sqrt{2D_{RR}t_{\rm sep}},
\end{equation}
from which we deduce
\begin{equation}
t_{\rm sep}=8 D_{RR}\left(\frac{\rho}{(\tau_1-\tau_2)\partial P/\partial R}\right)^2.
\end{equation}
If the size of population 1 particles is much bigger than that of population 2 particles, e.g. if the former are chondrules, or refractory inclusions, while the latter are matrix grains, we have:
\begin{eqnarray}
t_{\rm sep}\approx \Omega^{-1}\frac{\delta_R}{\textrm{St}^2}\left(\frac{v_K}{c_s}\right)^2
&=&2\:\mathrm{Myr}\:\frac{R_{\rm AU}^{1/2}}{S_R}\left(\frac{10^{-4}}{\textrm{St}}\right)\left(\frac{1\:\mathrm{km/s}}{c_s}\right)^2\nonumber\\
&=&\frac{t_{\rm drag}}{S_R}=\frac{t_{\rm vis}}{S^2\textrm{Sc}_R}
\label{tsep}
\end{eqnarray}
where $\textrm{St}$, $S$ and $S_R$ pertain to the larger particles
. 

  For $S\ll 1$, $t_{\rm sep}$ is longer than $t_{\rm vis}$ and thus ``decoherence'' cannot occur within the accretion timescale (which is necessarily bounded by that of the drift to the Sun). In this case, chondrules and dust grains remain coherent. Note that this is meant in a statistical sense, that is, the two populations \textit{as wholes} are not spatially resolved from one another ; of course, a particular chondrule and a particular dust grain floating nearby would not, on average, remain together unless they quickly coagulate.

  In contrast, for $S_R\gg 1$, $t_{\rm sep}$ is smaller than $t_{\rm drag}$, and judging from equation (\ref{tsep}), decoherence then seems unavoidable if chondrules and matrix do not agglomerate within $\ll$1 Myr in an evolved disk. That the accretion timescale was longer is evidenced by the broad chondrule age distribution in single meteorites \citep{Villeneuveetal2009}. In the ``inward transport'' simulations of \citet{Ciesla2009}, where particles are initially seeded at 20-30 AU, 
 significant decoherence between the $a=0.5\:\mathrm{mm}$ and the $a=5\:\mathrm{\mu m}$ particles can be observed for $\dot{M}=10^{-8}\:\rm M_\odot/yr$ (his Fig. 18), where the $S=1$ line lies at $R=4\:\mathrm{AU}$. 

  Matrix-chondrule complementarity in CCs 
 is evidence for coherence between chondrules and dust grains. Indeed, were chondrules unrelated to the matrix, that the rock resulting from their agglomeration is solar (e.g. in terms of the Mg/Si ratio) would seem too coincidental \citep[e.g.][]{HezelPalme2010}. Therefore, chondrules must have formed among dust grains now found in matrix, with exchange of chemical elements taking place between them during the chondrule-forming event while preserving a solar bulk composition. In addition, for this complementarity to have been maintained in CCs, both populations of objects must not have drifted apart until accretion: from the above discussion, this suggests that formation and accretion of CC components took place in regions where $S<1$.

  Following \citet{Cuzzietal2005}, we stress that the source of the components of a given CC needs \textit{not} be unique: If two source reservoirs A and B are solar in bulk composition, their respective contributions to a chondrite-forming region C will be solar too---since transport is unbiased with respect to component size (for $S\ll 1$)---and so will be the chondrite resulting from their mixing in region C. This chondrite will thus exhibit chondrule/matrix complementarity. Such an ``hybrid'' complementarity may be exemplified by CM chondrites, which \citet{Cassen1996} suggested to result from the mixing of CV/CO-like with CI-like materials. Hence, complementarity in one chondrite can be the result of \textit{several} distinct chondrule-forming events.


  As regards EORs, their composition was interpreted by \citet{LarimerAnders1970} as resulting from a loss of refractory, olivine-rich components from a starting material of solar composition. Lost components may include amoeboid olivine aggregates---a variety of refractory inclusions--- 
\citep{Ruzickaetal2012AOA} and chondrules, which tend to have supersolar Mg/Si in CCs \citep{HezelPalme2010}. Such a fractionation requires $S>1$, again consistent with our conjecture.

\subsection{Turbulent mixing and chondrule oxygen isotopes}
\label{Argument 4}
  Let us now focus on a single population of particles concentrated at some heliocentric distance at some initial time $t=0$. 

  For $S\ll 1$, diffusion will smear the particle distribution accross the whole range spanned in the inward transport. Indeed, using equation (\ref{tvis}), the eventual spread is of order:
\begin{equation}
\sqrt{D_{RR}t_{\rm vis}}=\frac{R}{\sqrt{\textrm{Sc}_R}}. 
\end{equation}
\citet{ClarkePringle1988} derived an exact solution illustrating this in the case of a steady disk with $\Sigma\propto R^{-2}$ and constant $\textrm{Sc}_R$, for perfectly coupled particles
: The particle distribution in terms of heliocentric distances is a log-normal, which, expressed as a function of $\mathrm{ln}R$, peaks at $R=R_0\exp{(-3\nu(R_0)t/2R_0)}$ (with $R_0$ the initial heliocentric distance) and has standard deviation $\sigma_{\rm lnR}=2\sqrt{\nu(R_0)t/\textrm{Sc}_RR_0^2}$.
 For $t\sim t_{\rm vis}$, the latter is of order unity, implying smearing of the distribution over a lengthscale comparable to $R$ by then.

  In contrast, for $S\gg 1$, drag-induced inward drift overwhelms diffusion. Using equation (\ref{tdrag}), we have\footnote{The spread $\Delta R$ is bounded by this value not only because of time, but also because $|v_{\rm drift}|$ generally increases with $R$, as noted by \citet{YoudinShu2002}. By incurring a quicker drift of the outer wing of the distribution compared to its inner wing, this tends to squeeze it. The balance with diffusion, $D_{RR}/\Delta R\sim\Delta R \partial |v_{\rm drift}|/\partial R$, indeed yields the value given in equation (\ref{spread limit}).}:
\begin{equation}
\sqrt{D_{RR}t_{\rm drag}}=\frac{R}{\sqrt{S_R}}
\label{spread limit}
\end{equation}

  Thus, a reservoir with $S\gg 1$ will not have received significant contributions having \textit{diffused} from further than this. 
Hence, for a given size, mixing between regions 
separated by more than $R/\sqrt{S_R}$ is insignificant. 


  This may explain why the oxygen isotopic composition in chondrules exhibits less variability in individual meteorites for EORs than for CCs
, if we have $S>1$ for EORs and $S<1$ for CCs. 
It is possible that the differences between the three ordinary chondrite groups---a subset of the EORs---are due to differential radial drift from a single reservoir, as there are systematic chondrule size differences between the groups \citep{Clayton2003,Zandaetal2006}, again consistent with $S>1$ for them.

\section{Implications}
\label{Implications}


\subsection{The value of $S$ in steady disks}
\label{S steady disks}

To get a sense of the value of $S$ and its variations in disks, we focus here on the stationary disk model (see Section \ref{Disk modeling}).

  In order to estimate $S$, we will need a temperature prescription. Using equations (\ref{steady mass accretion rate}) and (\ref{Tmid})
, we have:
\begin{equation}
T=\mathrm{max}\bigg[\left(\frac{3}{128\pi^2}\frac{\kappa m}{\sigma_{\rm SB}k_B\langle\alpha\rangle_\rho}\dot{M}^2\Omega^3\right)^{1/5}, 280\:\mathrm{K}\:f_T R_{\rm AU}^{-1/2}  \bigg],
\label{T steady}
\end{equation}
with $\kappa$ the 
 opacity, $R_{\rm AU}\equiv R/(1\:\mathrm{AU})$ and $f_T$ a dimensionless constant. The irradiation temperature is after \citet{Hayashi1981}
 but more realistic irradiation models \citep[e.g.][]{ChiangGoldreich1997} should not yield too different power law exponents.

\begin{figure}
\resizebox{\hsize}{!}{
\includegraphics{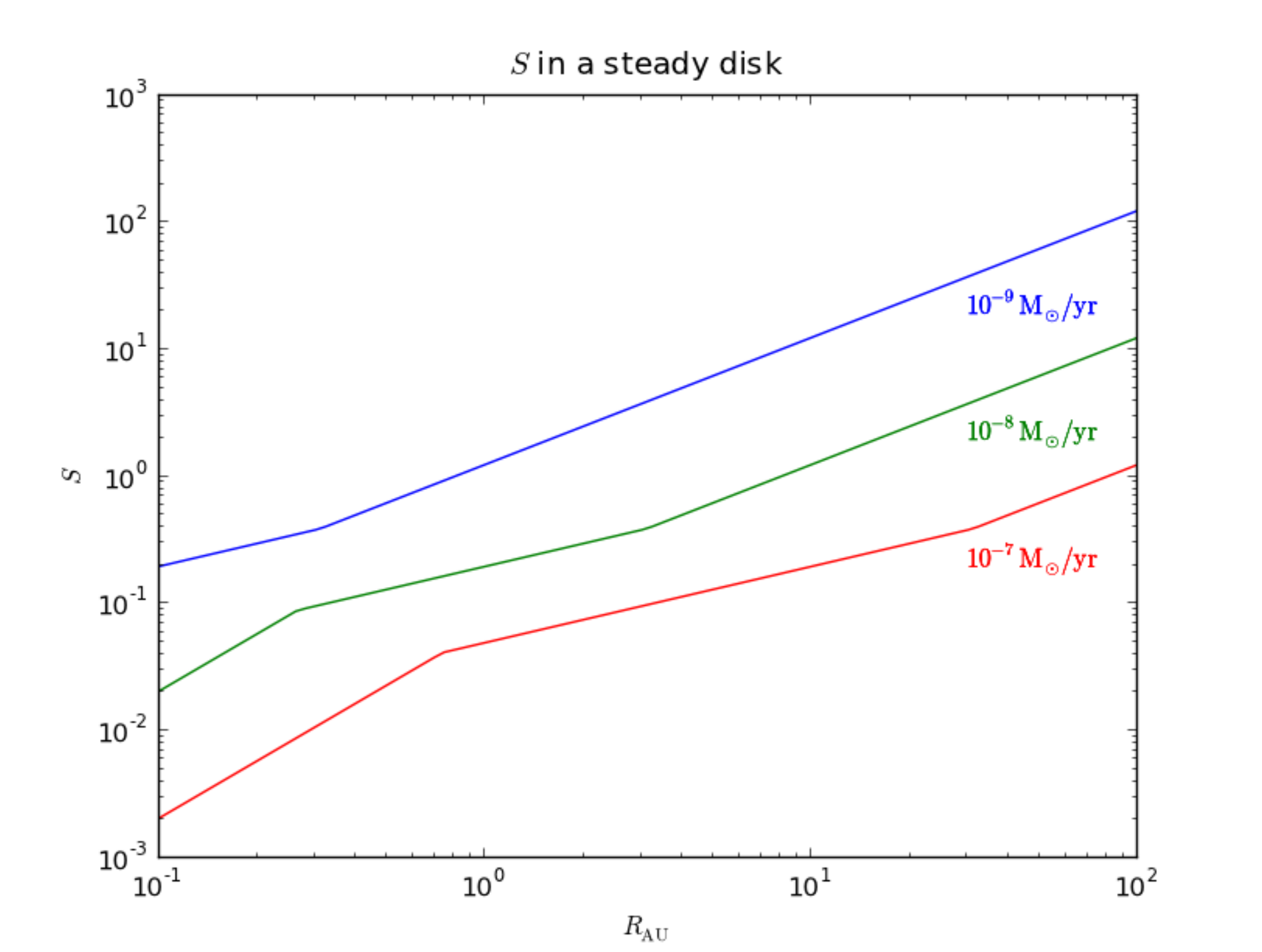}
}
\caption{Plot of the value of $S$ for chondrule-sized bodies ($\rho_sa = 1\:\rm kg/m^2$) in a steady disk, as a function of $R_{\rm AU}$, for three values of the mass accretion rate. 
Same parameters as Figure \ref{diffusion}. To crudely mimick the drop of opacity as solids evaporate, we limit the viscous temperature to 1500 K, hence the first break in the curves (except for $\dot{M}=10^{-9}\:\rm M_\odot\cdot yr^{-1}$). $S$ increases with heliocentric distance (and with decreasing mass accretion rate). The second break reflects a change in temperature regime from viscous heating- to irradiation-dominated.}
\label{plot S steady}
\end{figure}

\begin{figure}
\resizebox{\hsize}{!}{
\includegraphics{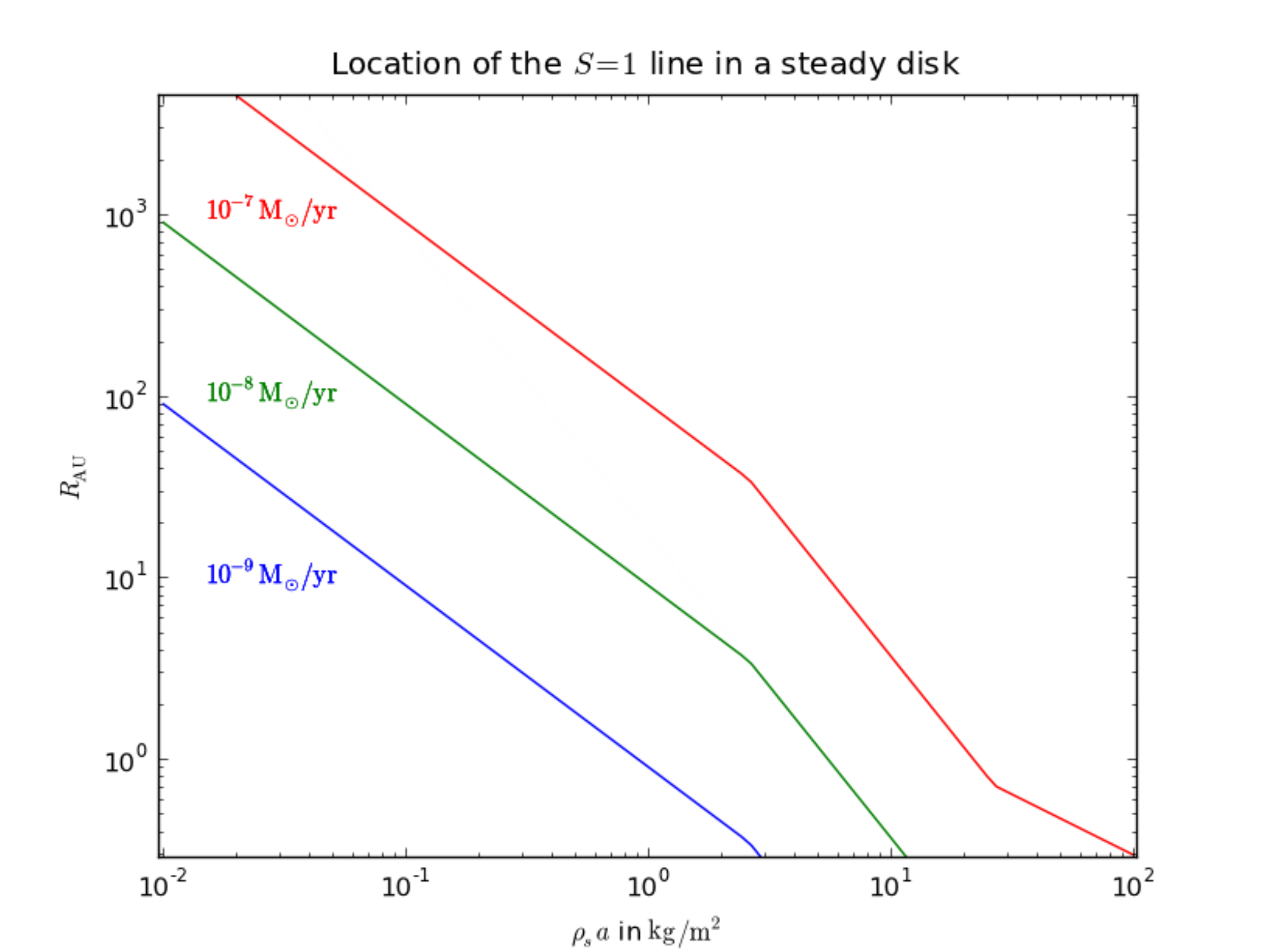}
}
\caption{Plot of location of the $S=1$ line as a function of $\rho_sa$ (recall that chondrule-sized bodies have $\rho_sa=1\:\rm kg/m^2$). Same parameters as Figure \ref{plot S steady}. As discussed in Section \ref{Argument 1}, this also corresponds to the maximum range of (significant) outward transport and decreases as expected with increasing size.}
\label{plot S1 steady}
\end{figure}

We have: 
\begin{equation}
\overline{S}\equiv\frac{\textrm{St}}{\langle\alpha\rangle_\rho}=\frac{3\pi^2}{2}\frac{\rho_sac_s^2}{\dot{M}\Omega}
\label{S steady}
\end{equation}
Importantly, explicit reference to $\alpha$ has vanished---even in the viscous heating-dominated regime, the dependence of the temperature on $\langle\alpha\rangle_\rho$ is weak, see equation (\ref{T steady})---and is replaced by the mass accretion rate, which is an astronomical observable. This means that $S$ can be evaluated without knowledge of the level of turbulence and its spatial variations in the disk. From equation (\ref{S steady}), one sees that $\overline{S}$ is an increasing function of $R$ and of time (that is, decreasing $\dot{M}$), as would have been the case in a constant $\alpha$ disk (excluding the viscous expansion phase), where $S\propto \Sigma^{-1}$, and which we expect on that basis to be a fairly general behavior. Numerically, we have: 
\begin{eqnarray}
\overline{S}=\left(\frac{\rho_sa}{1\:\mathrm{kg/m^2}}\right)\mathrm{max}\bigg[0.19\left(\frac{R_{\rm AU}}{\dot{M}_{-8}}\right)^{3/5}\left(\frac{10^{-3}}{\langle\alpha\rangle_\rho}\frac{\kappa}{5\:\mathrm{cm^2/g}}\right)^{1/5},\nonumber\\0.12f_T\frac{R_{\rm AU}}{\dot{M}_{-8}}\bigg],
\label{S steady numerical}
\end{eqnarray}
 with $\dot{M}_{-8}\equiv\dot{M}/(\rm 10^{-8}\:M_{\odot}\cdot yr^{-1})$
. This is plotted in Figure \ref{plot S steady}. We note that as the steady disk solution likely overestimates $\Sigma$ of real disks far from the Sun
, this likely underestimates $\overline{S}$ there. 
The heliocentric distance where $\overline{S}=1$ (plotted in Figure \ref{plot S1 steady}) is
\begin{eqnarray}
R_{\rm AU} = \dot{M}_{-8}\:\mathrm{min}\bigg[17\left(\frac{1\:\mathrm{kg/m^2}}{\rho_sa}\right)^{5/3}\left(\frac{\langle\alpha\rangle_\rho}{10^{-3}}\frac{5\:\mathrm{cm^2/g}}{\kappa}\right)^{1/3},\nonumber\\9\left(\frac{1\:\mathrm{kg/m^2}}{f_T\rho_sa}\right)\bigg].
\label{S=1}
\end{eqnarray}

\subsection{Chondrite diversity: Space or time?}
\label{time}


  Given our above expectation that $S$ should be an increasing function of heliocentric distance and time, our conjecture that CCs formed with $S<1$ while EORs formed with $S>1$ leaves two possiblities: either CCs formed closer to the Sun than EORs at some given epoch, or CC formation predated that of EORs, or some intermediate solution. We note that in both cases, the fact that EORs are more $^{16}$O-poor than CCs would be qualitatively consistent with models of O isotopic evolution involving isotopic exchange with $^{16}$O-poor water from the outer disk (e.g. CO self-shielding, see \citealt{Youngetal2008}):  such evolution would indeed have begun in the outer disk and then spread to the inner disk\footnote{However, samples returned from comet Wild 2 by the Stardust mission have oxygen isotopic signatures similar to CCs \citep{Oglioreetal2012}, although comets are thought to have accreted in the outer solar system
. This would be compatible only in case of an early accretion or a reduced isotopic exchange with the purported $^{16}$O-poor water in the outer disk compared to the inner disk.}. 

  The spatial alternative is unlikely, for the following reasons: 
The heliocentric distribution of asteroid taxonomic classes suggests that EOR parent bodies (presumably S-type asteroids and related classes) dominate the inner part of the main belt, while those of CCs (C-type asteroids and related classes) dominate its outer part \citep[e.g.][]{Burbineetal2008}. 
Also, many CCs display evidence of aqueous alteration and thus original presence of ice---stable far enough from the Sun--- on their parent body
, while evidence for water in EORs is comparatively marginal
. This runs counter to the spatial ordering that the conjecture on $S$ would suggest if time was not a significant factor.  

\begin{figure}\resizebox{\hsize}{!}{
\includegraphics{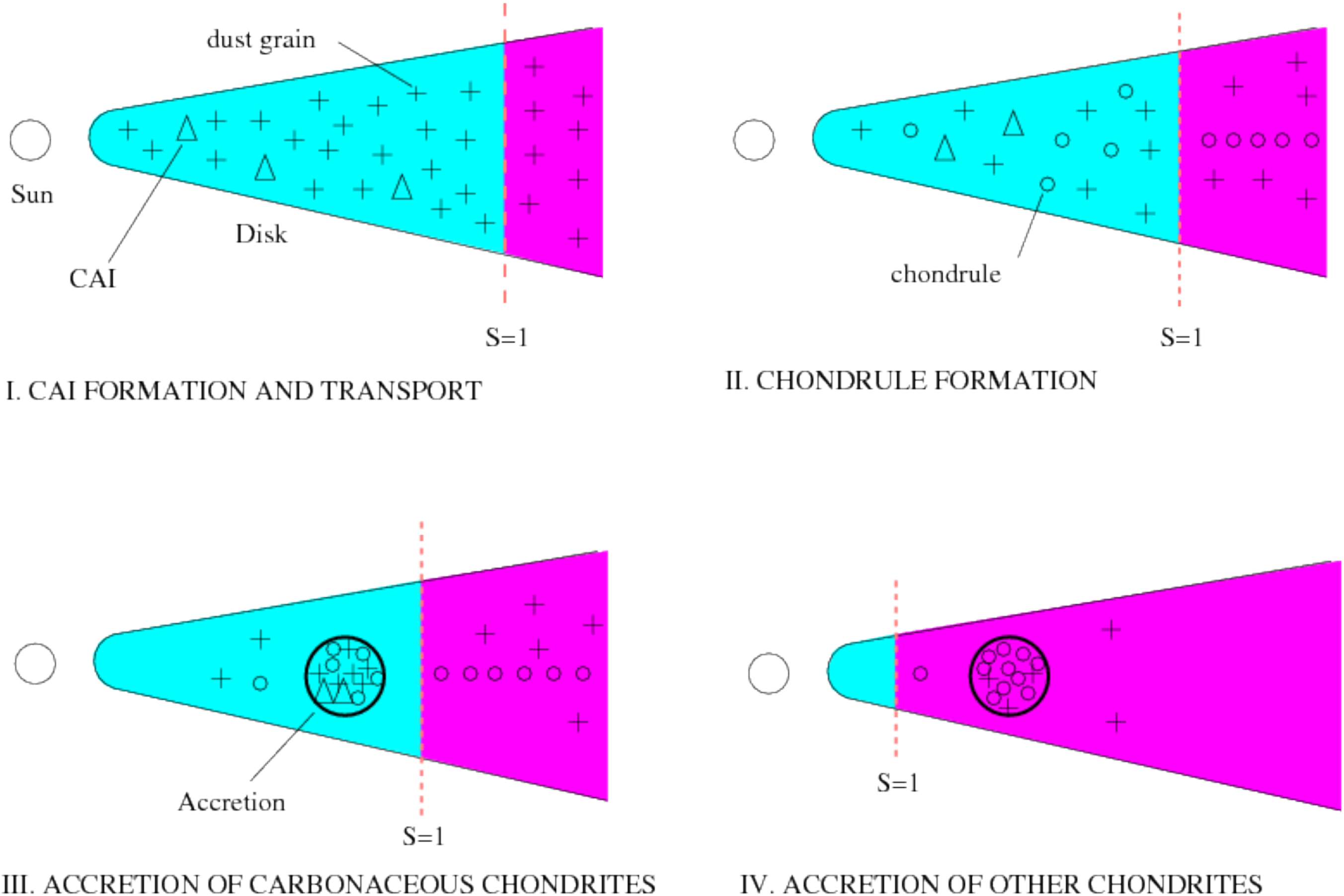}}
\caption{Cartoon of the proposed scenario. We show four schematic epochs (which may actually partly overlap): an early epoch where CAI are formed and transported (epoch ``I''; it would correspond to stages A-C of Fig. 1 of \citealt{Jacquetetal2011a}), the epoch of chondrule formation (epoch ``II''), the epochs of accretion of carbonaceous chondrites (epoch ``III'') and non-carbonaceous (``EOR'') chondrites (epoch ``IV''). We shade in blue those regions where $S<1$ and in pink those where $S>1$. A dashed line marks the position where $S=1$ and drifts inward as the disk evolves.  
 Millimeter-sized solids produced in the $S<1$ region are not efficiently transported to the $S>1$ region.}
\label{FigcartoonS}
\end{figure}

We thus suggest that \textit{CCs accreted earlier than EORs on average}, as also advocated by \citet{Cuzzietal2003} and \citet{Chambers2006}. This leads to the scenario sketched in Figure \ref{FigcartoonS}. It is worth pointing out that this \textit{chronological} interpretation could also account for the \textit{spatial} zoning of the main belt via an accretion efficiency effect. Indeed, the planetesimal formation rate may be taken to scale roughly like $\Sigma\Omega$ \citep[e.g.][]{Weidenschilling2004}
. For evolved phases of the disk where the surface density is low, this could have had a nonnegligible effect---bearing in mind that meteorite parent bodies probably formed over a wider range of heliocentric distances than the present-day main belt \citep[e.g.][]{Walshetal2011}---, accounting for preferential accretion of EOR parent bodies in the inner solar system.

  Our chronology is certainly consistent with the greater abundance of CAIs and amoeboid olivine aggregates (contemporaneous with the former, e.g. \citealt{SugiuraKrot2007}) in CCs than in EORs \citep{Cuzzietal2003,Chambers2006}, since these are the oldest solar system solids. 
 As to chondrules, age ranges seem as yet similar in CCs and EORs (1-3 Myr after CAI formation; \citealt{Villeneuveetal2009})
. This would not exclude, however, that they were \textit{accreted} earlier in CCs than in EORs.

  At early times, $^{26}$Al was a significant heat source in planetesimals and  many of them likely underwent differentiation \citep{SandersTaylor2005}. Inasmuch as materials precursor to many differentiated meteorites should thus have accreted as $S<1$, they would be expected to exhibit affinities to known CCs. Oxygen isotopes and depletion in volatile elements \citep[e.g.][]{WankeDreibus1986} 
 would be consistent with such a parentage \citep{MeibomClark1999} but other stable isotope data exclude a complete identification \citep{Warren2011b}. The CCs we observe today could have escaped extensive thermal processing e.g. because of the original presence of water \citep{Ghoshetal2006}, or as part of the unmelted periphery of their parent body as suggested by paleomagnetism measurements \citep{Weissetal2009}.

\subsection{Coexistence of CAIs and chondrules}
  We now investigate possible genetic relationships between CAIs and chondrules/matrix in carbonaceous chondrites. 

  We first emphasize that the contribution of the ``CAI factory'' to CCs cannot be reduced to the CAIs themselves, for the following two reasons: (i) There is no likely conceivable barrier for the condensation of most of the major rock-forming elements 
 after CAI formation. 
 Most condensable matter initially in the ``CAI factory'' must hence have eventually condensed, e.g. heterogeneously on the first condensates. 
 (ii) There is no \textit{dynamical} reason that CAIs should be transported outward preferentially to the remainder of the CAI factory matter. Actually, gas drag would rather lead to the \textit{opposite} effect
, but if $S\ll 1$ there, as we propose, there should be limited decoherence between CAIs and the remaining matter.

  It is noteworthy that then, incomplete condensation would not be ascribable to ``gas loss''  in the form originally hypothesized by \citet{WassonChou1974} because gas would have remained coherent with the solids. This leaves kinetic effects as a plausible alternative explanation. 

  From simple mass balance considerations \citep{Grossman2010}, CAIs likely constituted no more than $\sim$8 \% of the condensed matter from the CAI factory (assuming a water- and organics-free CI composition). This means that the few percent of CAIs reported in CCs imply that the ``CAI factory'' contribution was likely a few \textit{tens} of percent! Thus \textit{CAIs should not be thought of as a population foreign to the host chondrite} and introduced only at the time of accretion. The fact that the Al/Si ratio in CCs, while supersolar, becomes subsolar if the CAIs are (mentally) subtracted \citep{Hezeletal2008}, corroborates that there was some coherence between CAIs and the other components. 
This would imply that CAIs coexisted with chondrule precursors
, then chondrules, and some of them must have witnessed chondrule-forming events, as supported e.g. by the occurrence of chondrules with relict CAIs 
\citep{Krotetal2009}. 

  That CAIs and the surrounding matter remained coherent is not at variance with the fact that their environment evolved to match that in which chondrules formed. Initial isotopic heterogeneities would have been averaged out and mixing with matter from the outer disk or infalling on the disk would have diluted systematic isotopic anomalies. For example, the excesses in $^{54}$Cr and $^{46,50}$Ti measured in CAIs \citep{Trinquieretal2009} could have undergone dilution with isotopically differentiated meteorite-like matter to lead to the observed whole-rock composition of CCs. This is in the range suggested by CAI abundances.

\section{Summary and conclusions}
\label{Conclusion}
  We have considered a standard 
 turbulent disk. The turbulence parameter $\alpha$ was assumed not to vary by more than a factor of several in the vertical direction, except perhaps for surficial active layers (in presence of a dead zone), and Schmidt numbers were assumed to be of order unity (although we kept them explicit throughout). We have ignored possible large-scale nonaxisymmetric structures and the influence of planets. 

  In this class of disk models, we have derived the continuity equation for a population of grains in the tight coupling limit (that is, a dimensionless stopping time $\textrm{St}\ll 1$), in order to study their dynamics until agglomeration in $\gtrsim$ 1 cm objects. We have only considered transport of grains within the disk and ignored the effects of feedback of solids on the gas and that of accretion/fragmentation processes on the disk-scale transport of millimeter-sized bodies. A possible accretion bias against specific chondrite components was deemed irrelevant to explain the composition spectrum of chondrites although this issue would deserve more investigations.

  The dynamics of solid particles subject to gas drag are found to be essentially controlled by the ``gas-solid decoupling parameter'' $S\equiv \textrm{St}/\alpha$. For $S\ll 1$, particles essentially follow gas motions and are well-mixed vertically, while for $S\gg 1$, they settle to the midplane and drift inward faster than the gas. We find that a one-dimensional model is as yet most appropriate to evaluate net radial transport.

  We have attempted to estimate $S$ for the millimeter-sized chondrite components at the time of accretion for the different chondrites. We have conjectured that $S<1$ for carbonaceous chondrites (CCs) and $S>1$ for non-carbonaceous chondrites (EORs), and have put forward four arguments:

(i) Since particles produced close to the Sun cannot be efficiently transported beyond the $S=1$ line, the presence of calcium-aluminum-rich inclusions (CAIs) in CCs argues for $S<1$ for them.

(ii) The large chondrule-to-matrix ratio in EORs is consistent with preferential concentration of chondrules at the midplane as a result of settling, which requires $S>1$ for them.

(iii) Evidence for coherence between components of different grain size in CCs suggests $S<1$, whereas Mg/Si fractionation in EORs relative to solar abundances implies $S>1$.

(iv) The smaller range of oxygen isotopic compositions of chondrules in individual EORs compared to CCs could result from limited radial mixing in a $S>1$ environment. 

    $S$ has been evaluated for steady disks 
 and is found to be an increasing function of time and heliocentric distance. The above conjecture would thus favor earlier accretion of CCs compared to EORs, for which we listed some cosmochemical and planetological supporting evidence. Another implication is that CAIs belong to the same reservoir as part of the chondrule precursors for CCs but significant mixing between different reservoirs must have occurred before chondrule formation. 

  In this study, we have not attempted to introduce new physics, and we have recalled above some possibly relevant physical ingredients that we have ignored. Instead, we have tried to base our reasonings on as generic a disk as possible, in view of the existing literature on the transport of solids. The point of the present paper is that the simplest disk and radial transport models appear to lead to a robust conclusion about the space-time ordering of chondrite accretion. 
Our treatment of chondrites has been similarly generic: In being content with a first-order division of chondrites in two super-clans, irrespective of the specificities of the groups that they encompass, we have dealt in all rigor with schematic end-members. In particular, there is no reason to doubt that the transition from the $S<1$ to the $S>1$ regime was gradual and possibly some chondrite groups, e.g. those showing mixed affinities to CCs and EORs in terms of the criteria of Section 2 (e.g. CR chondrites, Kakangari-like chondrites...) 
might be representative of accretion at $S\sim 1$. 

\section*{Acknowledgement}
We thank Fred Ciesla and an anonymous referee for their thorough reviews that led to substantial improvements in the clarity and focus of the manuscript. Financial support from the Physique et Chimie du Milieu Interstellaire (PCMI) program and the Programme National de Plan\'{e}tologie (PNP) is gratefully acknowledged.

\begin{appendix}
\section{Vertical thermal structure of an optically thick disk region due to viscous dissipation}
In the optically thick limit (for $\Sigma > 1\:\rm g/cm^2$) and assuming local dissipation of turbulence, the vertical temperature profile is governed by the transfer equation \citep{Terquem2008}
\begin{equation}
\frac{\partial}{\partial z}\left(-\frac{4\sigma_{\rm SB}}{3\rho\kappa}\frac{\partial T^4}{\partial z}\right)=\frac{9}{4}\rho\alpha c_s^2\Omega,
\end{equation}
for timescales longer than the thermal timescale $(\Omega\alpha)^{-1}$. If we assume that 
$\kappa$ is vertically constant, this can be rewritten as
\begin{equation}
\frac{\partial^2T^4}{\partial\Sigma'^2}=-\frac{27\Omega\kappa}{16\sigma_{\rm SB}}\alpha c_s^2,
\label{transfer Sigma'}
\end{equation}
with $\Sigma'(z)\equiv \int_0^z\rho(z')\mathrm{d}z'$, 
 which, assuming that $\alpha c_s^2$ is vertically constant, may be integrated as:
\begin{equation}
T=T_{\rm mid}\left(1-\left(1-\left(\frac{T_{\rm irr}}{T_{\rm mid}}\right)^4\right)\left(\frac{2\Sigma'}{\Sigma}\right)^2\right)^{1/4},
\label{temperature profile}
\end{equation}
with $T_{\rm mid}$ the midplane temperature and $T_{\rm irr}$ the ``surface'' temperature\footnote{Or rather, the temperature where the diffusion approximation breaks down, which is below the ``superheated dust layer'' of \citet{ChiangGoldreich1997}
. Technically, this temperature is given by the implicit boundary condition of \citet{PapaloizouTerquem1999}, but in our limit of high optical thickness, only the irradiation contribution from above this height is important in equation (\ref{quartic sum}).}, both being related by:
\begin{equation}
T_{\rm mid}^4=T_{\rm irr}^4+\frac{27}{128\sigma_{\rm SB}}\kappa\Sigma^2\Omega\alpha c_s^2
\label{quartic sum}
\end{equation}
The solution may be approximated as:
\begin{equation}
T_{\rm mid}=\mathrm{max}\bigg[\left(\frac{27k_B}{128m\sigma_{\rm SB}}\kappa\Omega\Sigma^2\alpha\right)^{1/3}  ,T_{\rm irr}\bigg].
\label{Tmid}
\end{equation}

  Equation (\ref{temperature profile}) justifies the isothermal approximation over most of the column. Even if $T_{\rm irr}\ll T_{\rm mid}$, the temperature is within 20 \% of $T_{\rm mid}$ for 80 \% of the column. Had we 
given a temperature dependence to $\kappa$
, we would not 
 have obtained a significantly different $T_{\rm mid}$. In the irradiation-dominated regime, isothermality is also satisfied in the disk interior (between the ``superheated'' dust layers of \citealt{ChiangGoldreich1997}). 
While we have assumed that $\alpha c_s^2$ was vertically constant, variations of a factor of a few (notwithstanding the possibility of layered accretion) would also have little effect on $T_{\rm mid}$ because of the high power with which it appears in the equation.

\end{appendix}

\bibliographystyle{icarus}
\bibliography{bibliography} 

\begin{thebibliography}{}

\bibitem[{Amelin} {\em et~al.}(2010){Amelin}, {Kaltenbach}, {Iizuka},
  {Stirling}, {Ireland}, {Petaev}, and {Jacobsen}]{Amelinetal2010}
{Amelin}, Y., A.~{Kaltenbach}, T.~{Iizuka}, C.~H. {Stirling}, T.~R. {Ireland},
  M.~{Petaev},\ and S.~B. {Jacobsen} 2010.
\newblock {U-Pb chronology of the Solar System's oldest solids with variable
  $^{238}$U/$^{235}$U}.
\newblock {\em Earth and Planetary Science Letters\/}~{\bf 300}, 343--350.

\bibitem[{Anders}(1964){Anders}]{Anders1964}
{Anders}, E. 1964.
\newblock {Origin, age, and composition of meteorites}.
\newblock {\em Space Science Review\/}~{\bf 3}, 583--714.

\bibitem[{Bai} and {Stone}(2011){Bai} and {Stone}]{BaiStone2011}
{Bai}, X.-N.,\ and J.~M. {Stone} 2011.
\newblock {Effect of Ambipolar Diffusion on the Nonlinear Evolution of
  Magnetorotational Instability in Weakly Ionized Disks}.
\newblock {\em ApJ\/}~{\bf 736}, 144.

\bibitem[{Balbus}(2011){Balbus}]{Balbus2009}
{Balbus}, S.~A. 2011.
\newblock {\em {Magnetohydrodynamics of Protostellar Disks}}, pp.\  237--282.
\newblock University of Chicago Press.

\bibitem[{Balbus} and {Hawley}(1998){Balbus} and {Hawley}]{BalbusHawley1998}
{Balbus}, S.~A.,\ and J.~F. {Hawley} 1998.
\newblock {Instability, turbulence, and enhanced transport in accretion disks}.
\newblock {\em Reviews of Modern Physics\/}~{\bf 70}, 1--53.

\bibitem[{Balbus} and {Papaloizou}(1999){Balbus} and
  {Papaloizou}]{BalbusPapaloizou1999}
{Balbus}, S.~A.,\ and J.~C.~B. {Papaloizou} 1999.
\newblock {On the Dynamical Foundations of alpha Disks}.
\newblock {\em ApJ\/}~{\bf 521}, 650--658.

\bibitem[{Bland} {\em et~al.}(2005){Bland}, {Alard}, {Benedix}, {Kearsley},
  {Menzies}, {Watt}, and {Rogers}]{Blandetal2005}
{Bland}, P.~A., O.~{Alard}, G.~K. {Benedix}, A.~T. {Kearsley}, O.~N. {Menzies},
  L.~E. {Watt},\ and N.~W. {Rogers} 2005.
\newblock {Volatile fractionation in the early solar system and
  chondrule/matrix complementarity}.
\newblock {\em Proceedings of the National Academy of Science\/}~{\bf 102},
  13755--13760.

\bibitem[{Bockel{\'e}e-Morvan} {\em et~al.}(2002){Bockel{\'e}e-Morvan},
  {Gautier}, {Hersant}, {Hur{\'e}}, and {Robert}]{BockeleeMorvanetal2002}
{Bockel{\'e}e-Morvan}, D., D.~{Gautier}, F.~{Hersant}, J.~{Hur{\'e}},\ and
  F.~{Robert} 2002.
\newblock {Turbulent radial mixing in the solar nebula as the source of
  crystalline silicates in comets.}
\newblock {\em A\&A\/}~{\bf 384}, 1107--1118.

\bibitem[{Bouvier} and {Wadhwa}(2010){Bouvier} and {Wadhwa}]{BouvierWadhwa2010}
{Bouvier}, A.,\ and M.~{Wadhwa} 2010.
\newblock The age of the solar system redefined by the oldest pb-pb age of a
  meteoritic inclusion.
\newblock {\em Nature geoscience\/}~{\bf 3}, 637--641.

\bibitem[{Brearley} and {Jones}(1998){Brearley} and {Jones}]{BrearleyJones1998}
{Brearley}, A.,\ and A.~{Jones} 1998.
\newblock {\em Planetary Materials}, Chapter~3, pp.\  3--1--3--398.
\newblock MSA.

\bibitem[{Burbine} {\em et~al.}(2008){Burbine}, {Rivkin}, {Noble},
  {Moth\'{e}-Diniz}, {Bottke}, {McCoy}, and A.]{Burbineetal2008}
{Burbine}, T.~H., A.~S. {Rivkin}, S.~K. {Noble}, T.~{Moth\'{e}-Diniz},
  W.~{Bottke}, D.~M. {McCoy}, T.~J.,\ and T.~C. A. 2008.
\newblock {\em in Oxygen in the Solar System}, Chapter~12, pp.\  273--343.
\newblock MSA.

\bibitem[{Campbell} {\em et~al.}(2005){Campbell}, {Zanda}, {Perron}, {Meibom},
  and {Petaev}]{Campbelletal2005}
{Campbell}, A.~J., B.~{Zanda}, C.~{Perron}, A.~{Meibom},\ and M.~I. {Petaev}
  2005.
\newblock {Origin and Thermal History of Fe-Ni Metal in Primitive Chondrites}.
\newblock In {A.~N.~Krot, E.~R.~D.~Scott, \& B.~Reipurth} (Ed.), {\em
  Chondrites and the Protoplanetary Disk}, Volume 341 of {\em Astronomical
  Society of the Pacific Conference Series}, pp.\  407--431.

\bibitem[{Cassen}(1996){Cassen}]{Cassen1996}
{Cassen}, P. 1996.
\newblock {Models for the fractionation of moderately volatile elements in the
  solar nebula}.
\newblock {\em Meteoritics and Planetary Science\/}~{\bf 31}, 793--806.

\bibitem[{Cassen}(2001){Cassen}]{Cassen2001}
{Cassen}, P. 2001.
\newblock {Nebular thermal evolution and the properties of primitive planetary
  materials}.
\newblock {\em Meteoritics and Planetary Science\/}~{\bf 36}, 671--700.

\bibitem[{Chambers}(2006){Chambers}]{Chambers2006}
{Chambers}, J. 2006.
\newblock {\em {Meteoritic Diversity and Planetesimal Formation}}, Chapter~30,
  pp.\  487--497.

\bibitem[{Chiang} and {Goldreich}(1997){Chiang} and
  {Goldreich}]{ChiangGoldreich1997}
{Chiang}, E.~I.,\ and P.~{Goldreich} 1997.
\newblock {Spectral Energy Distributions of T Tauri Stars with Passive
  Circumstellar Disks}.
\newblock {\em ApJ\/}~{\bf 490}, 368--376.

\bibitem[{Ciesla}(2008){Ciesla}]{Ciesla2008}
{Ciesla}, F.~J. 2008.
\newblock {Radial transport in the solar nebula: Implications for moderately
  volatile element depletions in chondritic meteorites}.
\newblock {\em Meteoritics and Planetary Science\/}~{\bf 43}, 639--655.

\bibitem[{Ciesla}(2009a){Ciesla}]{Ciesla2009b}
{Ciesla}, F.~J. 2009a.
\newblock {Dynamics of high-temperature materials delivered by jets to the
  outer solar nebula}.
\newblock {\em M\&PS\/}~{\bf 44}, 1663--1673.

\bibitem[{Ciesla}(2009b){Ciesla}]{Ciesla2009}
{Ciesla}, F.~J. 2009b.
\newblock {Two-dimensional transport of solids in viscous protoplanetary
  disks}.
\newblock {\em Icarus\/}~{\bf 200}, 655--671.

\bibitem[{Ciesla}(2010){Ciesla}]{Ciesla2010}
{Ciesla}, F.~J. 2010.
\newblock {The distributions and ages of refractory objects in the solar
  nebula}.
\newblock {\em Icarus\/}~{\bf 208}, 455--467.

\bibitem[{Ciesla} and {Cuzzi}(2006){Ciesla} and {Cuzzi}]{CieslaCuzzi2006}
{Ciesla}, F.~J.,\ and J.~N. {Cuzzi} 2006.
\newblock {The evolution of the water distribution in a viscous protoplanetary
  disk}.
\newblock {\em Icarus\/}~{\bf 181}, 178--204.

\bibitem[{Clarke} and {Pringle}(1988){Clarke} and {Pringle}]{ClarkePringle1988}
{Clarke}, C.~J.,\ and J.~E. {Pringle} 1988.
\newblock {The diffusion of contaminant through an accretion disc}.
\newblock {\em MNRAS\/}~{\bf 235}, 365--373.

\bibitem[{Clayton}(2003){Clayton}]{Clayton2003}
{Clayton}, R.~N. 2003.
\newblock {\em in Treatise of Geochemistry}, Chapter 1.06, pp.\  130--142.
\newblock Elsevier.

\bibitem[{Connolly} and {Desch}(2004){Connolly} and {Desch}]{ConnollyDesch2004}
{Connolly}, H.~C., Jr.,\ and S.~J. {Desch} 2004.
\newblock {On the origin of the ''kleine Kugelchen'' called Chondrules}.
\newblock {\em Chemie der Erde/Geochemistry\/}~{\bf 64}, 95--125.

\bibitem[{Cuzzi} {\em et~al.}(2005){Cuzzi}, {Ciesla}, {Petaev}, {Krot},
  {Scott}, and {Weidenschilling}]{Cuzzietal2005}
{Cuzzi}, J.~N., F.~J. {Ciesla}, M.~I. {Petaev}, A.~N. {Krot}, E.~R.~D.
  {Scott},\ and S.~J. {Weidenschilling} 2005.
\newblock {Nebula Evolution of Thermally Processed Solids: Reconciling Models
  and Meteorites}.
\newblock In {A.~N.~Krot, E.~R.~D.~Scott, \& B.~Reipurth} (Ed.), {\em
  Chondrites and the Protoplanetary Disk}, Volume 341 of {\em Astronomical
  Society of the Pacific Conference Series}, pp.\  732--773.

\bibitem[{Cuzzi} {\em et~al.}(2003){Cuzzi}, {Davis}, and
  {Dobrovolskis}]{Cuzzietal2003}
{Cuzzi}, J.~N., S.~S. {Davis},\ and A.~R. {Dobrovolskis} 2003.
\newblock {Blowing in the wind. II. Creation and redistribution of refractory
  inclusions in a turbulent protoplanetary nebula}.
\newblock {\em Icarus\/}~{\bf 166}, 385--402.

\bibitem[{Cuzzi} {\em et~al.}(1996){Cuzzi}, {Dobrovolskis}, and
  {Hogan}]{Cuzzietal1996}
{Cuzzi}, J.~N., A.~R. {Dobrovolskis},\ and R.~C. {Hogan} 1996.
\newblock {Turbulence, chondrules, and planetesimals.}
\newblock In {R.~Hewins, R.~Jones, \& E.~Scott} (Ed.), {\em Chondrules and the
  Protoplanetary Disk}, pp.\  35--43.

\bibitem[{Cuzzi} {\em et~al.}(2001){Cuzzi}, {Hogan}, {Paque}, and
  {Dobrovolskis}]{Cuzzietal2001}
{Cuzzi}, J.~N., R.~C. {Hogan}, J.~M. {Paque},\ and A.~R. {Dobrovolskis} 2001.
\newblock {Size-selective Concentration of Chondrules and Other Small Particles
  in Protoplanetary Nebula Turbulence}.
\newblock {\em ApJ\/}~{\bf 546}, 496--508.

\bibitem[{Cuzzi} and {Weidenschilling}(2006){Cuzzi} and
  {Weidenschilling}]{CuzziWeidenschilling2006}
{Cuzzi}, J.~N.,\ and S.~J. {Weidenschilling} 2006.
\newblock {\em {Particle-Gas Dynamics and Primary Accretion}}, pp.\  353--381.

\bibitem[{Desch}(2007){Desch}]{Desch2007}
{Desch}, S.~J. 2007.
\newblock {Mass Distribution and Planet Formation in the Solar Nebula}.
\newblock {\em ApJ\/}~{\bf 671}, 878--893.

\bibitem[{Dubrulle} and {Frisch}(1991){Dubrulle} and
  {Frisch}]{DubrulleFrisch1991}
{Dubrulle}, B.,\ and U.~{Frisch} 1991.
\newblock {Eddy viscosity of parity-invariant flow}.
\newblock {\em Phys. Rev. A\/}~{\bf 43}, 5355--5364.

\bibitem[{Dubrulle} {\em et~al.}(1995){Dubrulle}, {Morfill}, and
  {Sterzik}]{Dubrulleetal1995}
{Dubrulle}, B., G.~{Morfill},\ and M.~{Sterzik} 1995.
\newblock {The dust subdisk in the protoplanetary nebula}.
\newblock {\em Icarus\/}~{\bf 114}, 237--246.

\bibitem[{Dzyurkevich} {\em et~al.}(2010){Dzyurkevich}, {Flock}, {Turner},
  {Klahr}, and {Henning}]{Dzyurkevichetal2010}
{Dzyurkevich}, N., M.~{Flock}, N.~J. {Turner}, H.~{Klahr},\ and T.~{Henning}
  2010.
\newblock {Trapping solids at the inner edge of the dead zone: 3-D global MHD
  simulations}.
\newblock {\em A\&A\/}~{\bf 515}, A70.

\bibitem[Epstein(1924)Epstein]{Epstein1924}
Epstein, P.~S. 1924.
\newblock On the resistance experienced by spheres in their motion through
  gases.
\newblock {\em Phys. Rev.\/}~{\em 23\/}(6), 710--733.

\bibitem[{Fleming} and {Stone}(2003){Fleming} and {Stone}]{FlemingStone2003}
{Fleming}, T.,\ and J.~M. {Stone} 2003.
\newblock {Local Magnetohydrodynamic Models of Layered Accretion Disks}.
\newblock {\em ApJ\/}~{\bf 585}, 908--920.

\bibitem[{Fromang} {\em et~al.}(2011){Fromang}, {Lyra}, and
  {Masset}]{Fromangetal2011}
{Fromang}, S., W.~{Lyra},\ and F.~{Masset} 2011.
\newblock {Meridional circulation in turbulent protoplanetary disks}.
\newblock {\em Astronomy \& Astrophysics\/}~{\bf 534}, A107.

\bibitem[{Fromang} and {Papaloizou}(2006){Fromang} and
  {Papaloizou}]{FromangPapaloizou2006}
{Fromang}, S.,\ and J.~{Papaloizou} 2006.
\newblock {Dust settling in local simulations of turbulent protoplanetary
  disks}.
\newblock {\em A\&A\/}~{\bf 452}, 751--762.

\bibitem[{Gail}(2001){Gail}]{Gail2001}
{Gail}, H.-P. 2001.
\newblock {Radial mixing in protoplanetary accretion disks. I. Stationary disc
  models with annealing and carbon combustion}.
\newblock {\em Astronomy \& Astrophysics\/}~{\bf 378}, 192--213.

\bibitem[{Gammie}(1996){Gammie}]{Gammie1996}
{Gammie}, C.~F. 1996.
\newblock {Layered Accretion in T Tauri Disks}.
\newblock {\em ApJ\/}~{\bf 457}, 355--362.

\bibitem[{Ghosh} {\em et~al.}(2006){Ghosh}, {Weidenschilling}, {McSween}, and
  {Rubin}]{Ghoshetal2006}
{Ghosh}, A., S.~J. {Weidenschilling}, H.~Y. {McSween}, Jr.,\ and A.~{Rubin}
  2006.
\newblock {\em {Asteroidal Heating and Thermal Stratification of the Asteroidal
  Belt}}, pp.\  555--566.

\bibitem[{Gounelle} {\em et~al.}(2007){Gounelle}, {Young}, {Shahar}, {Tonui},
  and {Kearsley}]{Gounelleetal2007}
{Gounelle}, M., E.~D. {Young}, A.~{Shahar}, E.~{Tonui},\ and A.~{Kearsley}
  2007.
\newblock {Magnesium isotopic constraints on the origin of CB$_{b}$
  chondrites}.
\newblock {\em Earth and Planetary Science Letters\/}~{\bf 256}, 521--533.

\bibitem[{Grossman}(2010){Grossman}]{Grossman2010}
{Grossman}, L. 2010.
\newblock {Vapor-condensed phase processes in the early solar system}.
\newblock {\em M\&PS\/}~{\bf 45}, 7--20.

\bibitem[{Hartmann} {\em et~al.}(1998){Hartmann}, {Calvet}, {Gullbring}, and
  {D'Alessio}]{Hartmannetal1998}
{Hartmann}, L., N.~{Calvet}, E.~{Gullbring},\ and P.~{D'Alessio} 1998.
\newblock {Accretion and the Evolution of T Tauri Disks}.
\newblock {\em ApJ\/}~{\bf 495}, 385--400.

\bibitem[{Hayashi}(1981){Hayashi}]{Hayashi1981}
{Hayashi}, C. 1981.
\newblock {Structure of the Solar Nebula, Growth and Decay of Magnetic Fields
  and Effects of Magnetic and Turbulent Viscosities on the Nebula}.
\newblock {\em Progress of Theoretical Physics Supplement\/}~{\bf 70}, 35--53.

\bibitem[{Hezel} and {Palme}(2010){Hezel} and {Palme}]{HezelPalme2010}
{Hezel}, D.~C.,\ and H.~{Palme} 2010.
\newblock {The chemical relationship between chondrules and matrix and the
  chondrule matrix complementarity}.
\newblock {\em Earth and Planetary Science Letters\/}~{\bf 294}, 85--93.

\bibitem[{Hezel} {\em et~al.}(2008){Hezel}, {Russell}, {Ross}, and
  {Kearsley}]{Hezeletal2008}
{Hezel}, D.~C., S.~S. {Russell}, A.~J. {Ross},\ and A.~T. {Kearsley} 2008.
\newblock {Modal abundances of CAIs: Implications for bulk chondrite element
  abundances and fractionations}.
\newblock {\em M\&PS\/}~{\bf 43}, 1879--1894.

\bibitem[{Hughes} and {Armitage}(2010){Hughes} and
  {Armitage}]{HughesArmitage2010}
{Hughes}, A.~L.~H.,\ and P.~J. {Armitage} 2010.
\newblock {Particle Transport in Evolving Protoplanetary Disks: Implications
  for Results from Stardust}.
\newblock {\em ApJ\/}~{\bf 719}, 1633--1653.

\bibitem[{Hughes} and {Armitage}(2012){Hughes} and
  {Armitage}]{HughesArmitage2012}
{Hughes}, A.~L.~H.,\ and P.~J. {Armitage} 2012.
\newblock {Global variation of the dust-to-gas ratio in evolving protoplanetary
  discs}.
\newblock {\em ArXiv e-prints\/}.

\bibitem[{Ilgner} and {Nelson}(2008){Ilgner} and {Nelson}]{IlgnerNelson2008}
{Ilgner}, M.,\ and R.~P. {Nelson} 2008.
\newblock {Turbulent transport and its effect on the dead zone in
  protoplanetary discs}.
\newblock {\em Astronomy \& Astrophysics\/}~{\bf 483}, 815--830.

\bibitem[{Jacquet} {\em et~al.}(2011){Jacquet}, {Fromang}, and
  {Gounelle}]{Jacquetetal2011a}
{Jacquet}, E., S.~{Fromang},\ and M.~{Gounelle} 2011.
\newblock {Radial transport of refractory inclusions and their preservation in
  the dead zone}.
\newblock {\em A\&A\/}~{\bf 526}, L8.

\bibitem[{Johansen} {\em et~al.}(2006){Johansen}, {Klahr}, and
  {Mee}]{Johansenetal2006}
{Johansen}, A., H.~{Klahr},\ and A.~J. {Mee} 2006.
\newblock {Turbulent diffusion in protoplanetary discs: the effect of an
  imposed magnetic field}.
\newblock {\em MNRAS\/}~{\bf 370}, L71--L75.

\bibitem[{Jones} {\em et~al.}(2000){Jones}, {Lee}, {Connolly}, {Love}, and
  {Shang}]{Jonesetal2000}
{Jones}, R.~H., T.~{Lee}, H.~C. {Connolly}, Jr., S.~G. {Love},\ and H.~{Shang}
  2000.
\newblock {Formation of Chondrules and CAIs: Theory VS. Observation}.
\newblock {\em Protostars and Planets IV\/}, 927--962.

\bibitem[{Kallemeyn} {\em et~al.}(1996){Kallemeyn}, {Rubin}, and
  {Wasson}]{Kallemeynetal1996}
{Kallemeyn}, G.~W., A.~E. {Rubin},\ and J.~T. {Wasson} 1996.
\newblock {The compositional classification of chondrites: VII. The R chondrite
  group}.
\newblock {\em Geochimica et Cosmochimica Acta\/}~{\bf 60}, 2243--2256.

\bibitem[{King} and {King}(1978){King} and {King}]{KingKing1978}
{King}, T.~V.~V.,\ and E.~A. {King} 1978.
\newblock {Grain size and petrography of C2 and C3 carbonaceous chondrites}.
\newblock {\em Meteoritics\/}~{\bf 13}, 47--72.

\bibitem[{Krot} {\em et~al.}(2009){Krot}, {Amelin}, {Bland}, {Ciesla},
  {Connelly}, {Davis}, {Huss}, {Hutcheon}, {Makide}, {Nagashima}, {Nyquist},
  {Russell}, {Scott}, {Thrane}, {Yurimoto}, and {Yin}]{Krotetal2009}
{Krot}, A.~N., Y.~{Amelin}, P.~{Bland}, F.~J. {Ciesla}, J.~{Connelly}, A.~M.
  {Davis}, G.~R. {Huss}, I.~D. {Hutcheon}, K.~{Makide}, K.~{Nagashima}, L.~E.
  {Nyquist}, S.~S. {Russell}, E.~R.~D. {Scott}, K.~{Thrane}, H.~{Yurimoto},\
  and Q.-Z. {Yin} 2009.
\newblock {Origin and chronology of chondritic components: A review}.
\newblock {\em Geochimica et Cosmochimica Acta\/}~{\bf 73}, 4963--4997.

\bibitem[{Krot} {\em et~al.}(2005){Krot}, {Amelin}, {Cassen}, and
  {Meibom}]{Krotetal2005}
{Krot}, A.~N., Y.~{Amelin}, P.~{Cassen},\ and A.~{Meibom} 2005.
\newblock {Young chondrules in CB chondrites from a giant impact in the early
  Solar System}.
\newblock {\em Nature\/}~{\bf 436}, 989--992.

\bibitem[{Kuebler} {\em et~al.}(1999){Kuebler}, {McSween}, {Carlson}, and
  {Hirsch}]{KueblerMcSween1999}
{Kuebler}, K.~E., H.~Y. {McSween}, W.~D. {Carlson},\ and D.~{Hirsch} 1999.
\newblock {Sizes and Masses of Chondrules and Metal-Troilite Grains in Ordinary
  Chondrites: Possible Implications for Nebular Sorting}.
\newblock {\em Icarus\/}~{\bf 141}, 96--106.

\bibitem[{Larimer} and {Anders}(1970){Larimer} and {Anders}]{LarimerAnders1970}
{Larimer}, J.~W.,\ and E.~{Anders} 1970.
\newblock {Chemical fractionations in meteorites--III. Major element
  fractionations in chondrites}.
\newblock {\em Geochimica et Cosmochimica Acta\/}~{\bf 34}, 367--387.

\bibitem[Lathrop {\em et~al.}(1992)Lathrop, Fineberg, and
  Swinney]{Lathropetal1992}
Lathrop, D.~P., J.~Fineberg,\ and H.~L. Swinney 1992.
\newblock Transition to shear-driven turbulence in couette-taylor flow.
\newblock {\em Phys. Rev. A\/}~{\bf 46}, 6390--6405.

\bibitem[{Lesur} and {Longaretti}(2007){Lesur} and
  {Longaretti}]{LesurLongaretti2007}
{Lesur}, G.,\ and P.-Y. {Longaretti} 2007.
\newblock {Impact of dimensionless numbers on the efficiency of
  magnetorotational instability induced turbulent transport}.
\newblock {\em MNRAS\/}~{\bf 378}, 1471--1480.

\bibitem[{MacPherson}(2005){MacPherson}]{MacPherson2005}
{MacPherson}, G.~J. 2005.
\newblock {\em {Treatise on Geochemistry}}, Chapter 1.08, pp.\  201--246.
\newblock Elsevier B.

\bibitem[{May} {\em et~al.}(1999){May}, {Russell}, and {Grady}]{Mayetal1999}
{May}, C., S.~S. {Russell},\ and M.~M. {Grady} 1999.
\newblock {Analysis of Chondrule and CAI Size and Abundance in CO3 and CV3
  Chondrites: A Preliminary Study}.
\newblock In {\em Lunar and Planetary Institute Science Conference Abstracts},
  Volume~30, pp.\  1688.

\bibitem[{Meibom} and {Clark}(1999){Meibom} and {Clark}]{MeibomClark1999}
{Meibom}, A.,\ and B.~E. {Clark} 1999.
\newblock {Invited review: Evidence for the insignificance of ordinary
  chondritic material in the asteroid belt}.
\newblock {\em M\&PS\/}~{\bf 34}, 7--24.

\bibitem[{Miller} and {Stone}(2000){Miller} and {Stone}]{MillerStone2000}
{Miller}, K.~A.,\ and J.~M. {Stone} 2000.
\newblock {The Formation and Structure of a Strongly Magnetized Corona above a
  Weakly Magnetized Accretion Disk}.
\newblock {\em The Astrophysical Journal\/}~{\bf 534}, 398--419.

\bibitem[{Ogliore} {\em et~al.}(2011){Ogliore}, {Huss}, {Nagashima},
  {Butterworth}, {Gainsforth}, {Stodolna}, {Westphal}, {Joswiak}, and
  {Tyliszczak}]{Oglioreetal2012}
{Ogliore}, R.~C., G.~R. {Huss}, K.~{Nagashima}, A.~L. {Butterworth},
  Z.~{Gainsforth}, J.~{Stodolna}, A.~J. {Westphal}, D.~{Joswiak},\ and
  T.~{Tyliszczak} 2011.
\newblock {Incorporation of a Late-forming Chondrule into Comet Wild 2}.
\newblock {\em ArXiv e-prints\/}.

\bibitem[{Oishi} and {Mac Low}(2009){Oishi} and {Mac Low}]{OishiMcLow2009}
{Oishi}, J.~S.,\ and M.-M. {Mac Low} 2009.
\newblock {On Hydrodynamic Motions in Dead Zones}.
\newblock {\em ApJ\/}~{\bf 704}, 1239--1250.

\bibitem[{Palme} and {Jones}(2005){Palme} and {Jones}]{PalmeJones2005}
{Palme}, H.,\ and A.~{Jones} 2005.
\newblock {\em {Meteorites, Comets and Planets: Treatise on Geochemistry,
  Volume 1}}, Chapter {Solar System Abundances of the Elements}, pp.\ ~41.
\newblock Elsevier.

\bibitem[{Papaloizou} and {Terquem}(1999){Papaloizou} and
  {Terquem}]{PapaloizouTerquem1999}
{Papaloizou}, J.~C.~B.,\ and C.~{Terquem} 1999.
\newblock {Critical Protoplanetary Core Masses in Protoplanetary Disks and the
  Formation of Short-Period Giant Planets}.
\newblock {\em ApJ\/}~{\bf 521}, 823--838.

\bibitem[{Prinn}(1990){Prinn}]{Prinn1990}
{Prinn}, R.~G. 1990.
\newblock {On neglect of nonlinear momentum terms in solar nebula accretion
  disk models}.
\newblock {\em ApJ\/}~{\bf 348}, 725--729.

\bibitem[{Rubin}(2000){Rubin}]{Rubin2000}
{Rubin}, A.~E. 2000.
\newblock {Petrologic, geochemical and experimental constraints on models of
  chondrule formation}.
\newblock {\em Earth Science Reviews\/}~{\bf 50}, 3--27.

\bibitem[{Rubin}(2010){Rubin}]{Rubin2010}
{Rubin}, A.~E. 2010.
\newblock {Physical properties of chondrules in different chondrite groups:
  Implications for multiple melting events in dusty environments}.
\newblock {\em Geochimica et Cosmochimica Acta\/}~{\bf 74}, 4807--4828.

\bibitem[{Rubin}(2011){Rubin}]{Rubin2011}
{Rubin}, A.~E. 2011.
\newblock {Origin of the differences in refractory-lithophile-element
  abundances among chondrite groups}.
\newblock {\em Icarus\/}~{\bf 213}, 547--558.

\bibitem[{Ruzicka} {\em et~al.}(2012){Ruzicka}, {Floss}, and
  {Hutson}]{Ruzickaetal2012AOA}
{Ruzicka}, A., C.~{Floss},\ and M.~{Hutson} 2012.
\newblock {Amoeboid olivine aggregates (AOAs) in the Efremovka, Leoville and
  Vigarano (CV3) chondrites: A record of condensate evolution in the solar
  nebula}.
\newblock {\em Geochimica et Cosmochimica Acta\/}~{\bf 79}, 79--105.

\bibitem[{Sanders} and {Taylor}(2005){Sanders} and {Taylor}]{SandersTaylor2005}
{Sanders}, I.~S.,\ and G.~J. {Taylor} 2005.
\newblock {Implications of aluminum-26 in Nebular Dust: Formation of Chondrules
  by Disruption of Molten Planetesimals}.
\newblock In {A.~N.~Krot, E.~R.~D.~Scott, \& B.~Reipurth} (Ed.), {\em
  Chondrites and the Protoplanetary Disk}, Volume 341 of {\em Astronomical
  Society of the Pacific Conference Series}, pp.\  915--932.

\bibitem[{Schr{\"a}pler} and {Henning}(2004){Schr{\"a}pler} and
  {Henning}]{SchraeplerHenning2004}
{Schr{\"a}pler}, R.,\ and T.~{Henning} 2004.
\newblock {Dust Diffusion, Sedimentation, and Gravitational Instabilities in
  Protoplanetary Disks}.
\newblock {\em ApJ\/}~{\bf 614}, 960--978.

\bibitem[{Scott} and {Krot}(2003){Scott} and {Krot}]{ScottKrot2003}
{Scott}, E.~R.~D.,\ and A.~N. {Krot} 2003.
\newblock {Chondrites and their Components}.
\newblock {\em Treatise on Geochemistry\/}~{\bf 1}, 143--200.

\bibitem[{Shu} {\em et~al.}(2001){Shu}, {Shang}, {Gounelle}, {Glassgold}, and
  {Lee}]{Shuetal2001}
{Shu}, F.~H., H.~{Shang}, M.~{Gounelle}, A.~E. {Glassgold},\ and T.~{Lee} 2001.
\newblock {The Origin of Chondrules and Refractory Inclusions in Chondritic
  Meteorites}.
\newblock {\em ApJ\/}~{\bf 548}, 1029--1050.

\bibitem[{Sugiura} and {Krot}(2007){Sugiura} and {Krot}]{SugiuraKrot2007}
{Sugiura}, N.,\ and A.~N. {Krot} 2007.
\newblock {26Al-26Mg systematics of Ca-Al-rich inclusions, amoeboid olivine
  aggregates, and chondrules from the ungrouped carbonaceous chondrite Acfer
  094}.
\newblock {\em Meteoritics and Planetary Science\/}~{\bf 42}, 1183--1195.

\bibitem[{Takeuchi} and {Lin}(2002){Takeuchi} and {Lin}]{TakeuchiLin2002}
{Takeuchi}, T.,\ and D.~N.~C. {Lin} 2002.
\newblock {Radial Flow of Dust Particles in Accretion Disks}.
\newblock {\em ApJ\/}~{\bf 581}, 1344--1355.

\bibitem[{Terquem}(2008){Terquem}]{Terquem2008}
{Terquem}, C.~E.~J.~M.~L.~J. 2008.
\newblock {New Composite Models of Partially Ionized Protoplanetary Disks}.
\newblock {\em ApJ\/}~{\bf 689}, 532--538.

\bibitem[{Trinquier} {\em et~al.}(2009){Trinquier}, {Elliott}, {Ulfbeck},
  {Coath}, {Krot}, and {Bizzarro}]{Trinquieretal2009}
{Trinquier}, A., T.~{Elliott}, D.~{Ulfbeck}, C.~{Coath}, A.~N. {Krot},\ and
  M.~{Bizzarro} 2009.
\newblock {Origin of Nucleosynthetic Isotope Heterogeneity in the Solar
  Protoplanetary Disk}.
\newblock {\em Science\/}~{\bf 324}, 374--376.

\bibitem[{Turner} {\em et~al.}(2010){Turner}, {Carballido}, and
  {Sano}]{Turneretal2010}
{Turner}, N.~J., A.~{Carballido},\ and T.~{Sano} 2010.
\newblock {Dust Transport in Protostellar Disks Through Turbulence and
  Settling}.
\newblock {\em ApJ\/}~{\bf 708}, 188--201.

\bibitem[{Villeneuve} {\em et~al.}(2009){Villeneuve}, {Chaussidon}, and
  {Libourel}]{Villeneuveetal2009}
{Villeneuve}, J., M.~{Chaussidon},\ and G.~{Libourel} 2009.
\newblock Homogeneous distribution of aluminum-26 in the solar system from the
  magnesium isotopic composition of chondrules.
\newblock {\em Science\/}~{\bf 325}, 985--988.

\bibitem[{Walsh} {\em et~al.}(2011){Walsh}, {Morbidelli}, {Raymond}, {O'Brien},
  and {Mandell}]{Walshetal2011}
{Walsh}, K.~J., A.~{Morbidelli}, S.~N. {Raymond}, D.~P. {O'Brien},\ and A.~M.
  {Mandell} 2011.
\newblock {A low mass for Mars from Jupiter's early gas-driven migration}.
\newblock {\em Nature\/}~{\bf 475}, 206--209.

\bibitem[{W{\"a}nke} and {Dreibus}(1986){W{\"a}nke} and
  {Dreibus}]{WankeDreibus1986}
{W{\"a}nke}, H.,\ and G.~{Dreibus} 1986.
\newblock {Die chemische Zusammensetzung und Bildung der terrestrischen
  Planeten}.
\newblock {\em Mitteilungen der Astronomischen Gesellschaft Hamburg\/}~{\bf
  65}, 9--24.

\bibitem[{Warren}(2011){Warren}]{Warren2011b}
{Warren}, P.~H. 2011.
\newblock {Stable-isotopic anomalies and the accretionary assemblage of the
  Earth and Mars: A subordinate role for carbonaceous chondrites}.
\newblock {\em Earth and Planetary Science Letters\/}~{\bf 311}, 93--100.

\bibitem[{Wasson} and {Chou}(1974){Wasson} and {Chou}]{WassonChou1974}
{Wasson}, J.~T.,\ and C.~{Chou} 1974.
\newblock {Fractionation of Moderately Volatile Elements in Ordinary
  Chondrites}.
\newblock {\em Meteoritics\/}~{\bf 9}, 69--84.

\bibitem[{Wehrstedt} and {Gail}(2002){Wehrstedt} and {Gail}]{WehrstedtGail2002}
{Wehrstedt}, M.,\ and H.~{Gail} 2002.
\newblock {Radial mixing in protoplanetary accretion disks. II. Time dependent
  disk models with annealing and carbon combustion}.
\newblock {\em A\&A\/}~{\bf 385}, 181--204.

\bibitem[{Weidenschilling}(2004){Weidenschilling}]{Weidenschilling2004}
{Weidenschilling}, S.~J. 2004.
\newblock {\em {From icy grains to comets}}, pp.\  97--104.

\bibitem[{Weiss} {\em et~al.}(2009){Weiss}, {Gattacceca}, {Stanley},
  {Rochette}, and {Christensen}]{Weissetal2009}
{Weiss}, B.~P., J.~{Gattacceca}, S.~{Stanley}, P.~{Rochette},\ and U.~R.
  {Christensen} 2009.
\newblock {Paleomagnetic Records of Meteorites and Early Planetesimal
  Differentiation}.
\newblock {\em Space Science Reviews\/}, 123--172.

\bibitem[{Youdin} and {Goodman}(2005){Youdin} and {Goodman}]{YoudinGoodman2005}
{Youdin}, A.~N.,\ and J.~{Goodman} 2005.
\newblock {Streaming Instabilities in Protoplanetary Disks}.
\newblock {\em ApJ\/}~{\bf 620}, 459--469.

\bibitem[{Youdin} and {Lithwick}(2007){Youdin} and
  {Lithwick}]{YoudinLithwick2007}
{Youdin}, A.~N.,\ and Y.~{Lithwick} 2007.
\newblock {Particle stirring in turbulent gas disks: Including orbital
  oscillations}.
\newblock {\em Icarus\/}~{\bf 192}, 588--604.

\bibitem[{Youdin} and {Shu}(2002){Youdin} and {Shu}]{YoudinShu2002}
{Youdin}, A.~N.,\ and F.~H. {Shu} 2002.
\newblock {Planetesimal Formation by Gravitational Instability}.
\newblock {\em ApJ\/}~{\bf 580}, 494--505.

\bibitem[{Young} {\em et~al.}(2008){Young}, {Kuramoto}, {Marcus}, {Yurimoto},
  and {Jacobsen}]{Youngetal2008}
{Young}, E.~D., K.~{Kuramoto}, R.~A. {Marcus}, H.~{Yurimoto},\ and S.~B.
  {Jacobsen} 2008.
\newblock {\em in Oxygen in the Solar System}, Chapter~9, pp.\  187--218.
\newblock MSA.

\bibitem[{Yurimoto} {\em et~al.}(2008){Yurimoto}, {Krot}, {Choi}, {Aléon},
  {Kunihiro}, and {Brearley}]{Yurimotoetal2008}
{Yurimoto}, H., A.~N. {Krot}, B.-G. {Choi}, J.~{Aléon}, T.~{Kunihiro},\ and
  A.~J. {Brearley} 2008.
\newblock {\em in Oxygen in the Solar System}, Chapter~8, pp.\  141--186.
\newblock MSA.

\bibitem[{Zanda} {\em et~al.}(2006){Zanda}, {Hewins}, {Bourot-Denise}, {Bland},
  and {Albar{\`e}de}]{Zandaetal2006}
{Zanda}, B., R.~H. {Hewins}, M.~{Bourot-Denise}, P.~A. {Bland},\ and
  F.~{Albar{\`e}de} 2006.
\newblock {Formation of solar nebula reservoirs by mixing chondritic
  components}.
\newblock {\em Earth and Planetary Science Letters\/}~{\bf 248}, 650--660.

\bibitem[{Zanda} {\em et~al.}(2012){Zanda}, {Humayun}, and
  {Hewins}]{Zandaetal2012}
{Zanda}, B., M.~{Humayun},\ and R.~H. {Hewins} 2012.
\newblock {Chemical Composition of Matrix and Chondrules in Carbonaceous
  Chondrites: Implications for Disk Transport}.
\newblock In {\em Lunar and Planetary Institute Science Conference Abstracts},
  Volume~43 of {\em Lunar and Planetary Institute Science Conference
  Abstracts}, pp.\  2413.

\bibitem[{Zhu} {\em et~al.}(2010){Zhu}, {Hartmann}, and {Gammie}]{Zhuetal2010b}
{Zhu}, Z., L.~{Hartmann},\ and C.~{Gammie} 2010.
\newblock {Long-term Evolution of Protostellar and Protoplanetary Disks. II.
  Layered Accretion with Infall}.
\newblock {\em ApJ\/}~{\bf 713}, 1143--1158.

\end{thebibliography}
\end{document}